\documentclass[sigconf]{acmart}
\acmConference[ISSTA 2022]{ACM SIGSOFT International Symposium on Software Testing and Analysis}{18-22 July, 2022}{Daejeon, South Korea}

\usepackage{xspace}
\usepackage{listings}
\usepackage{array}
\usepackage{algorithmic}
\usepackage{multirow}
\usepackage{tablefootnote}
\usepackage{hyperref}
\usepackage{cleveref}
\usepackage{booktabs}
\usepackage{tabularx}
\usepackage{tcolorbox}
\usepackage[table,x11names]{xcolor}
\usepackage[ruled,linesnumbered,noend]{algorithm2e}

\newcolumntype{Y}{>{\centering\arraybackslash}X}

\newcommand{\name}{GLAD\xspace}
\newcommand{\dfj}{Defects4J\xspace}
\newcommand{\ift}{\texttt{if}\xspace}

\lstdefinelanguage{diff}{
  basicstyle=\ttfamily\small,
  morecomment=[f][\color{gray}]{@@},
  morecomment=[l][\color{Green4}]{+\ },
  morecomment=[l][\color{red}]{-\ },
}

\AtBeginDocument{%
  \providecommand\BibTeX{{%
    \normalfont B\kern-0.5em{\scshape i\kern-0.25em b}\kern-0.8em\TeX}}}

\setcopyright{acmcopyright}
\copyrightyear{2022}
\acmYear{2022}
\acmDOI{10.1145/1122445.1122456}

\begin{document}

\title{\name: Neural Predicate Synthesis to Repair Omission Faults}

\author{Sungmin Kang}
\email{sungmin.kang@kaist.ac.kr}
\affiliation{%
  \institution{KAIST}
  \city{Daejeon}
  \country{Republic of Korea}
  \postcode{34141}
}

\author{Shin Yoo}
\email{shin.yoo@kaist.ac.kr}
\affiliation{%
  \institution{KAIST}
  \city{Daejeon}
  \country{Republic of Korea}
  \postcode{34141}
}

\renewcommand{\shortauthors}{Kang and Yoo}

\begin{abstract}
  Existing template and learning-based APR tools have successfully found 
  patches for many benchmark faults. However, our analysis of existing results 
  shows that omission faults pose a significant challenge to these techniques. 
  For template based approaches, omission faults provide no location to apply 
  templates to; for learning based approaches that formulate repair as Neural 
  Machine Translation (NMT), omission faults similarly do not provide the 
  faulty code to translate. To address these issues, we propose \name, a novel 
  learning-based repair technique that specifically targets if-clause synthesis. 
  \name does not require a faulty line as it is based on generative 
  Language Models (LMs) instead of machine translation; consequently, it can 
  repair omission faults. \name intelligently constrains the language model 
  using a type-based grammar. Further, it efficiently reduces the validation
  cost by performing dynamic ranking of candidate patches using a debugger. 
  Thanks to the shift from translation to synthesis, 
  \name is highly orthogonal to existing techniques: \name can correctly fix 16 \dfj v1.2 faults 
  that previous NMT-based techniques could not, while maintaining a reasonable runtime cost,
  underscoring its utility as an APR tool and potential to complement existing tools in practice.
  An inspection of the bugs that \name fixes reveals that \name can quickly generate
  expressions that would be challenging for other techniques.
\end{abstract}




\maketitle

\section{Introduction}
In Automated Program Repair (APR), one seeks to automatically fix faulty code given a specification of desired behavior (e.g. a logical constraint or a test suite~\cite{Gazolla2019as}). Researchers have attempted different approaches to tackle the problem; for example, SequenceR~\cite{Chen2019SequenceRSL} uses deep learning models trained on fix data to automatically find patches, while FixMiner~\cite{Koyuncu2020FixMinerMR} uses templates mined from human patches and applies them to fix faulty code. APR is an active research area, with more than 20 papers being published every year~\cite{Liu2020ot}. Given the large volume of APR research, it would be beneficial to look at the performance of our community as a whole. 

As many APR techniques are evaluated on the \dfj~\cite{Rene2014dj} dataset, we can assess
the collective performance of the APR community by analyzing which faults have been fixed
by any tool that deals with \dfj at the time of this writing. In short, 
we find that since the introduction of the \dfj dataset,
138 faults have been fixed overall. More importantly, our analysis reveals that the amount
of \textit{added characters} in the developer patch is strongly indicative of
APR success: while 91.7\% of faults that require fewer than 16 characters to be added are fixed, only 22.7\% of faults that require more than 32 characters to be added were ever fixed.

Inspecting the patches consisting of 16 to 63 added characters but have never been found by any APR tools, we find that a large portion contains
\textit{omission faults}, i.e. faults in which necessary code is missing. 
Omission faults pose a problem to both template-based and learning-based techniques. 
Template-based APR tools may fix only a restricted set of omission faults 
such as adding null checks, as there is no faulty code to apply templates to. 
Meanwhile, learning-based techniques use the neural machine \textit{translation} formulation, 
which necessitates a faulty statement to translate into a fixed statement, 
making omission faults difficult to fix. 
On the other hand, language models (LMs) are a natural solution to repairing omission faults, as they are trained with a generative loss~\cite{radford2018GPT1} and are consequently better suited to handle omission faults.

To this end, we introduce \name (Grammar-based LAnguage models with Debuggers)
which specializes in fixing \ift statement omission faults. 
To the best of our knowledge, \name is the first deep learning-based
technique that moves away from the translation paradigm to tackle omission faults. 
\name utilizes trained language models and type-based grammar rules to effectively synthesize
and modify predicates for \ift statements. 
Given an LM, \name guides it to generate predicates by placing
an \ift token at a given location, and queries the LM to synthesize natural sequences that can follow.
A well-trained LM will synthesize predicate sequences, given the presentation of the
\ift token. For efficient search, we introduce a variant of the standard beam search algorithm
that incorporates a type-based grammar and a vocabulary set.
Constraining the output of the LM allows \name to
generate significantly more valid expressions, enhancing its overall performance.

\name further narrows the space of potential patches by using a \textit{debugger},
which can both provide static information such as the legal members of an object, and 
dynamic information such as evaluation results of expressions, on the fly. By using the values of generated predicates on passing and failing tests, we can dynamically filter and rank potential patches without compiling or executing them, leading to efficient repair behavior. 

We perform empirical analysis to demonstrate the utility of \name.
First, we present the faults that \name fixed, showing the extent to
which \name has expanded the set of fixed faults in \dfj. Results show that \name
could fix 16 faults in \dfj v1.2 that previous learning-based techniques could not, and
overall we report fixing eight faults that were never fixed by any APR tool we examine. 
We verify that \name continues to generate correct patches in the bugs added by \dfj v2.0,
implying the versatility of \name's repair abilities.
We also find both the grammar-based beam search and the
debugger-based reranking make a significant contribution towards
reducing the space of patches to evaluate, showing that \name can combine these
information sources harmoniously to generate useful patches. An analysis on the runtime of 
\name reveals that it fixes the majority of bugs in under 30 minutes, confirming it is an effective repair tool; 
along with \name's orthogonal results, this also demonstrates how
\name could supplement other repair techniques to
expand the breadth of bugs that can be repaired using APR techniques in practice.

Qualitatively inspecting the bugs that \name uniquely fixes, we find that they
require the synthesis of complex expressions that are difficult to generate
using existing techniques, establishing the contribution of \name to the field of APR at large.
Further, these expressions are generated in less than 30 minutes, demonstrating
the efficiency of \name even when generating long predicates.

To summarize, our contribution is as follows:

\begin{itemize}
  \item We present an analysis of faults in the \dfj dataset that have been fixed
  by all published APR techniques we could identify, finding that the next level 
  of difficulty in APR lies in fixing omission faults;
  \item We introduce \name, which specializes in fixing \ift statement omission faults using novel technical components such as LMs (that can effectively generate omitted code), grammar rules (that can effectively constrain LM output), and a debugger (that can efficiently rank generated patches);
  \item We evaluate the repair performance of \name, showing that it is orthogonal to existing APR techniques, fixing 16 new faults relative to existing learning-based techniques and eight faults that no APR tool we survey has fixed, while managing to do this in a reasonable amount of time;
  \item We perform ablation studies to confirm that each component we add contributes towards successful repair;
  \item We make our tool publicly available\footnote{\url{https://anonymous.4open.science/r/neural-pred-synth-4816/README.md}}.
\end{itemize}

The organization of the paper is as follows. In \Cref{sec:rel_work}, we provide an overview of recent APR techniques, showing that many lack the ability to successfully deal with omission faults. Concurrent to this point, the data provided in \Cref{sec:apr_difficulty} shows that omission faults constitute the `next level' of APR difficulty that the APR community should overcome. In \Cref{sec:approach}, we detail \name, showing how it can specifically deal with \ift-statement omission faults. We pose the research questions in ~\Cref{sec:expr_details}, with results presented in \Cref{sec:results} showing that \name fixes 16 faults never fixed by any other learning-based APR tool, and is highly orthogonal to all APR techniques we examine. \Cref{sec:ttv} discusses threats to validity in our experiments, and \Cref{sec:conclusion} concludes.

\section{Related Work}
\label{sec:rel_work}

The academic context for our work is outlined in this section.

\subsection{Automated Program Repair}

Giving an overview of all program repair techniques is not within the scope of this paper; those interested may refer to Gazzola et al.~\cite{Gazolla2019as} for a broad array of techniques. Instead, we briefly describe techniques relevant to the purpose of \name.

\subsubsection{Template-based Techniques.} Since PAR~\cite{Kim2013par}, which introduced eight fix patterns derived from common developer patches, many new templates have been used to fix faults: AVATAR introduces templates learned from static analysis tools~\cite{Liu2019AVATARFS}, while FixMiner automatically mines patch templates from corpora of patch data~\cite{Koyuncu2020FixMinerMR}. TBar, among the strongest APR tools, collates templates used by existing APR tools and uses them to repair, resulting in a single tool fixing 74 faults\footnote{c.f. Liu et al.~\cite{Liu2019tbar}, Tab. 5} correctly~\cite{Liu2019tbar}. While template-based techniques have enjoyed strong performance, barring a few common patterns that have been captured by templates, such as adding null checkers (TBar FP2), template-based techniques are ill-suited in solving omission faults. This is why, although every possible output of the templates of TBar was evaluated on the correct position, not even plausible patches were made for omission faults such as Closure-15 of \dfj v1.2.

\subsubsection{Learning-based Program Repair}
Here, we focus on the deep learning (or NMT) based program repair techniques. Inspired by the success of NMT in the NLP field, Hata et al.~\cite{Hata2018LearningTG} and Chen et al.~\cite{Chen2019SequenceRSL} use the seq2seq~\cite{Sutskever2014SequenceTS} architecture, formulating repair as translation between faulty code and correct code. All work we are aware of maintains this formulation~\cite{Li2020DLFixCB,Chakraborty2020CODIT,Mashhadi2021BERTAPR,Zhu2021Recoder}; the current best-performing repair technique, Recoder~\cite{Zhu2021Recoder}, comes closer to synthesis by generating edits based on an edit grammar, but is still far from full-fledged synthesis as it only allows one identifier `placeholder' as part of its repair process. There is a fundamental difference between the NMT-based techniques and \name: NMT-based APR techniques structurally rely on existing faulty statements to generate patches, while \name only needs the code context leading to the faulty location.

\subsubsection{Condition Synthesis}
There is a significant body of work targeting condition synthesis; as condition synthesis often deals with omission faults, most of these techniques are not strictly template-based. ACS~\cite{Xiong2017acs} uses predicate switching and syntactic code search over a database to find fixing predicates; as such, ACS cannot fix condition omissions with a predicate that never appears in the code corpus, unlike \name. Nopol~\cite{DeMarco2014ar} and its successor Dynamoth~\cite{Durieux2016dynamoth} first base their predicate synthesis on oracle mining, then use an SMT solver and a debugger respectively to find the fixing predicate. The assumptions that Nopol/Dynamoth make are inaccurate at times, leading to significant test suite overfitting~\cite{Liu2020ot}. Finally, JAID~\cite{Chen2018JAID}/Restore~\cite{Xu2020Restore} use state abstraction to identify `suspicious program states'. Both are more restricted in the types of predicates they can generate than \name, which can search through a larger space efficiently thanks to language models.

\subsection{Language Models}
A statistical language model defines a probability distribution over the space of sentences. That is, for every possible sentence, a language model will provide the probability of such a sentence. Language models are trained so that the probability of a training corpus is maximized. Pretrained LMs are receiving a great amount of attention, thanks to their strong performance on few-shot learning~\cite{Brown2020GPT2}. LMs trained on source code are known to be able to measure naturalness~\cite{Hindle2012nat,Ray2016buggynat} and perform autocompletion tasks~\cite{Kim2021CodePF,chen2021codex}, so it is natural that they could perform the role of a synthesizer as well. We use byte-pair encoding (BPE) in this work, recently used in the software engineering context by Karampatsis et al.~\cite{Karampatsis2020bcbv}, to handle subtokenization. To the best of our knowledge we are the first to primarily use a language model to perform repair.

\subsection{Debuggers}
While \name is the first learning-based APR technique to employ debuggers, they have a history of use by APR tools. One early such work is Zeller~\cite{Zeller2002ic}, which uses a debugger (GDB) to isolate failure-relevant program states. Later, Galenson et al.~\cite{Galenson2014CodeHint} introduce CodeHint, which queries a debugger to identify available local variables, among others. CodeHint requires user-specified constraints to operate, making it difficult to compare with \name. Dynamoth~\cite{Durieux2016dynamoth} uses debuggers to generate and evaluate expressions as well.

\section{Analysis of Existing APR Results}
\label{sec:apr_difficulty}
We begin by investigating the state-of-the-art in APR as a whole: which faults 
in the \dfj dataset are reported to be correctly fixed by any APR tool? Following 
the criteria used by Liu et al.~\cite{Liu2020ot}, we survey papers 
in the community-maintained \textit{program-repair.org}, as well as in the 
living APR review maintained by Monperrus~\cite{monperrus2020livingreview}, to 
identify Java-targeting APR techniques. We apply the following criteria when 
gathering all faults fixed by the community:

\begin{itemize}
  \item As APR performance is often measured under controlled (i.e. perfect) fault localization~\cite{Liu2020ot,Lutellier2020cc,Li2020DLFixCB} to gauge repair performance without localization bias, when possible we use repair results under such conditions, either from the paper or from the results reported by Liu et al.~\cite{Liu2020ot}.
  \item If the repair performance under perfect fault localization is not available and the paper introducing the technique provides specific \dfj faults that were fixed, those are directly used as the results;
  \item We consider two different TBar versions, $TBar_p$~\cite{Liu2019tbar} and TBar-10k~\cite{Liu2020ot}, as the performance of TBar significantly changes based on the stopping criteria.
\end{itemize}

Performing this survey, we identify 46 tools that fix \dfj faults; among those, we could retrieve fault-level repair results for 40 tools. Each APR tool, along with their fault localization (FL) granularity, is reported in \Cref{tab:all_tools}. The source for these results can be found in the Appendix, Table 1. Overall, we find that there are 138 faults in \dfj that have been fixed at least once, with easy faults being fixed multiple times (e.g. the most fixed fault, Math-70, was fixed by 25 tools), while others are fixed only a few times (e.g. Math-71 was only fixed by SimFix). 

\begin{table}[ht]
  \footnotesize
  \centering
  \caption{All tools used for the state-of-the-art APR analysis. We could not find fault-level results for unused tools.\label{tab:all_tools}}
  \scalebox{0.9}{
  \begin{tabular}{p{0.19\linewidth}|>{\centering\arraybackslash}p{0.02\linewidth}|>{\raggedright\arraybackslash}p{0.66\linewidth}}
  \toprule
  FL & \# & Techniques \\
  \midrule
  Perfect-Statement & 23 & TBar-Time~\cite{Liu2019tbar}, TBar-10k~\cite{Liu2020ot}, CURE~\cite{Lutellier2021ca}, SequenceR~\cite{Chen2019SequenceRSL}, SimFix~\cite{Jiang2018ShapingPR}, FixMiner~\cite{Koyuncu2020FixMinerMR}, DLFix~\cite{Li2020DLFixCB}, ACS~\cite{Xiong2017acs}, Ratchet~\cite{Hata2018Ratchet}, CODIT~\cite{Chakraborty2020CODIT}, AVATAR~\cite{Liu2019AVATARFS}, CoCoNuT~\cite{Lutellier2020cc}, ARJA~\cite{Yuan2020ARJA}, jGenProg~\cite{Martinez2016ASTOR}, GenProg-A~\cite{Yuan2020ARJA}, jMutRepair~\cite{Martinez2016ASTOR}, kPAR~\cite{Liu2019kPAR}, RSRepair-A~\cite{Yuan2020ARJA}, jKali~\cite{Martinez2016ASTOR}, Kali-A~\cite{Yuan2020ARJA}, Dynamoth~\cite{Durieux2016dynamoth}, Nopol~\cite{DeMarco2014ar}, Cardumen~\cite{Martinez2018Cardumen} \\\hline
  Method-given & 4 & HDRepair~\cite{Le2016HistoryDP}, JAID~\cite{Chen2018JAID}, SketchFix~\cite{Hua2018SketchFix}, Restore~\cite{Xu2020Restore}\\\hline
  No perfect info, uses test spectra & 11 & Hercules~\cite{Saha2019HarnessingEF}, CapGen~\cite{Wen2018CapGen}, PAR~\cite{Kim2013par}, SOFix~\cite{Liu2018SOFix}, PraPR~\cite{Ghanbari2019PraPR}, ConFix~\cite{Kim2019ConFix}, VFix~\cite{Xu2019VFix}, iFixR~\cite{Koyuncu2019IFixR}, LSRepair~\cite{Liu2018LSRepair}, GenPat~\cite{Jiang2019GenPat}, Recoder~\cite{Zhu2021Recoder} \\\hline
  Others & 2 & ssFix~\cite{Xin2017ssFix}\tablefootnote{ssFix uses stack trace information.}, NPEFix~\cite{Cornu2015NPEFix}\tablefootnote{NPEFix has no fault localization.} \\\hline
  Unused in Study & 6 & Elixir~\cite{Saha2017Elixir}, DeepRepair~\cite{White2019DeepRepair} LoopRepair~\cite{Wang2019LoopFix}, xPAR~\cite{Le2016HistoryDP}, S3~\cite{Le2017S3}, VarFix~\cite{Wong2021VarFix}\\
  \bottomrule
  \end{tabular}
  }
\end{table}

We further obtain features for each fault based on the developer patch and analyze if there is a feature that predicts whether a fault will have been fixed by an APR tool at some point. If such a feature exists, we can reasonably say that the feature correlates with APR difficulty. We use the following features:

\begin{itemize}
  \item We use the \dfj-dissection dataset~\cite{Sobreira2018dissection} which provides 8 numeric features: the number of files/classes/methods/lines that are involved with the developer patch, the number of added/removed/modified lines, and the number of chunks.
  \item We use the \texttt{git diff} tool to obtain token-level diffs, and in turn derive characters added/removed features from them. As the diffs are token-level, the number of characters added/removed is always an overestimation.
\end{itemize}

\begin{figure}[h!]
  \centering
  \includegraphics[width=1.0\linewidth]{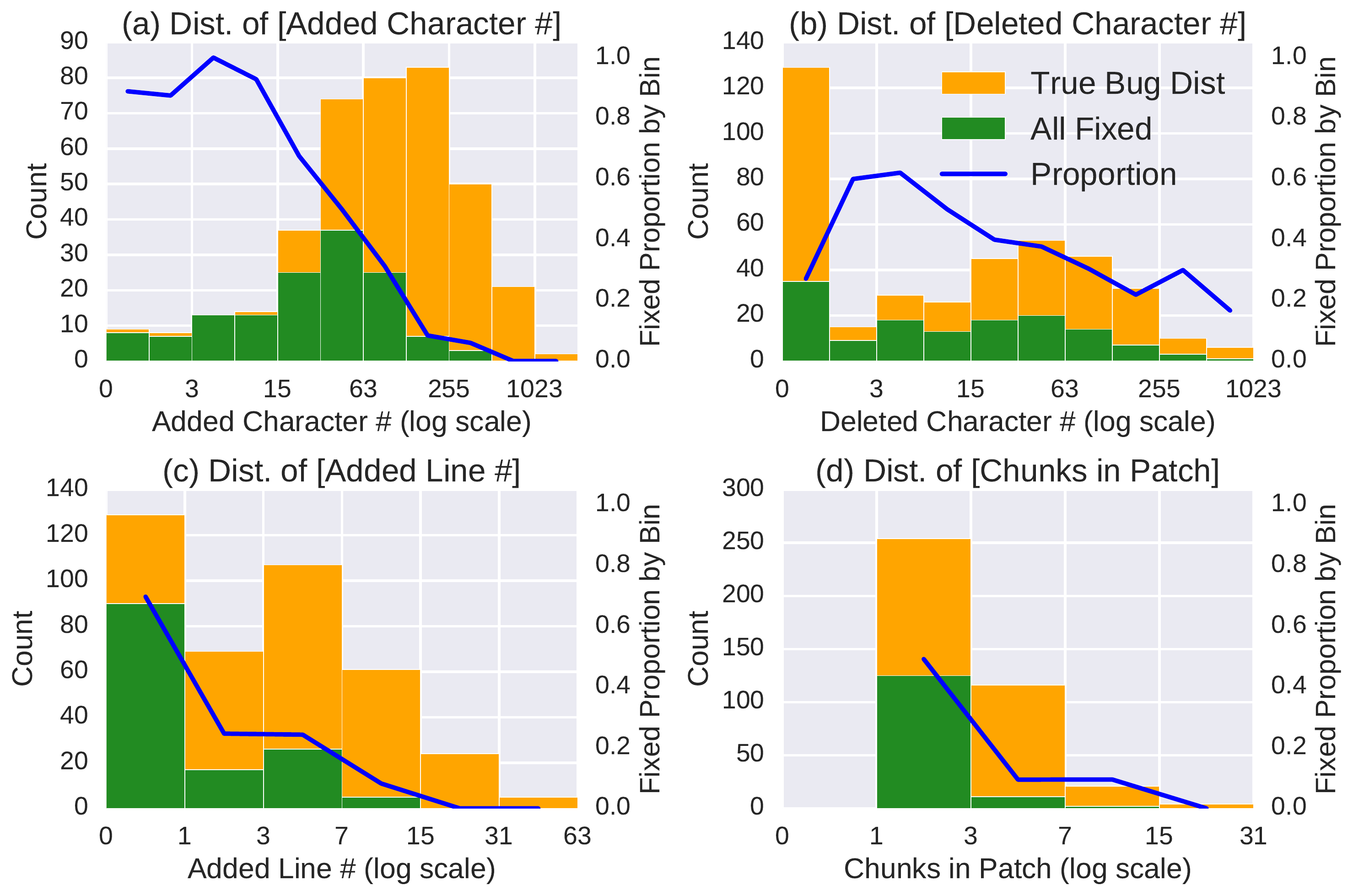}
  \caption{\dfj full fault distribution by feature, contrasted with the fixed fault distribution. The blue line represents the fixed fault proportion in a bin.}
  \label{fig:bug_dist}
\end{figure}

\noindent The raw data used in this section is made publicly available\footnote{\url{https://anonymous.4open.science/r/neural-pred-synth-4816/d4j_apr_results/fix_data.csv}}, so that the community can build upon it and perform other analyses.

Using these features, we plot the distribution difference between all faults in \dfj and those that have been fixed at least once. The results of this analysis are presented in \Cref{fig:bug_dist}. First, note that the amount of deletion (\Cref{fig:bug_dist}(b)) does not have a consistent relationship with the probability of a fault being patched (the blue line fluctuates), suggesting that deletion is not particularly relevant when accessing APR difficulty. However, the other three features have a clear negative relationship with patch probability. Following the fixed proportion (blue line) of \Cref{fig:bug_dist}(a), for faults whose patches contain a low number of added characters, the fixed proportion is highest of all features, indicating almost all have been fixed. As the added character number increases (and the graph moves to the right), the percentage of faults that have been fixed steeply decreases. Thus we suggest that the number of added characters is a reasonable measure of APR difficulty. This implies that \textbf{synthesis is the bottleneck of program repair}. As a result, the point where the fixed ratio decreases in \Cref{fig:bug_dist}(a) (15+ characters added) constitutes the `next level' of difficulty for APR.

\begin{table}[ht]
  \centering
  \caption{Unfixed `next level' faults in each fault category.\label{tab:bug_categories}}
  \scalebox{0.9}{
  \begin{tabular}{*{6}{c}}
  \toprule
  Fault Size & total & Weak om. & Strong. om. & Total om. & if om.\\\midrule
  16-31 & 11 & 4 (36.4\%) & 5 (45.5\%) & 8 (72.7\%) & 4 (36.4\%) \\
  32-63 & 37 & 9 (24.3\%) & 23 (62.2\%) & 31 (83.7\%) & 13 (35.1\%) \\\midrule
  Overall & 48 & 13 (27.1\%) & 28 (58.3\%) & 39 (81.3\%) & 17 (35.4\%) \\
  \bottomrule
  \end{tabular}
  }
\end{table}

What types of faults constitute the `next level'? We first define a few terms: a fault has `weak omission' if the developer patch contains a modification that only adds code to a certain statement. In theory, existing NMT or template-based techniques can fix weak omission faults, but the synthesis aspect can make fixing them  difficult. A fault has `strong omission' if a new statement is introduced. Strong omission faults are structurally challenging to fix for existing NMT-based techniques; depending on what the inserted code is, they can be difficult or impossible for existing template-based techniques as well. Finally, a fault has `if omission' if it has either weak or strong omission and the added code is related to an \ift predicate. We manually investigated all 48 unfixed faults within the 16-63 added character range. The results of this analysis, shown in \Cref{tab:bug_categories}, indicate that 81.3\% of all faults inspected have an omission aspect; the faults in which a novel statement is added constitute 58.3\% of unfixed faults in the range. Meanwhile the faults that have \ift-statement omission amount to 35.4\% (or 43.6\% of all omission faults), many of which require complex predicates that have resisted synthesis attempts until now. This is notable as 
omission faults, especially strong omission faults, are ill-handled by existing template-based and learning-based techniques, as noted in \Cref{sec:rel_work}. As the leading techniques in the field struggle in fixing omission faults, we believe a technique capable of efficiently complementing these tools represents a meaningful step forward in APR.

\begin{figure*}[h!]
  \centering
  \includegraphics[width=0.9\textwidth]{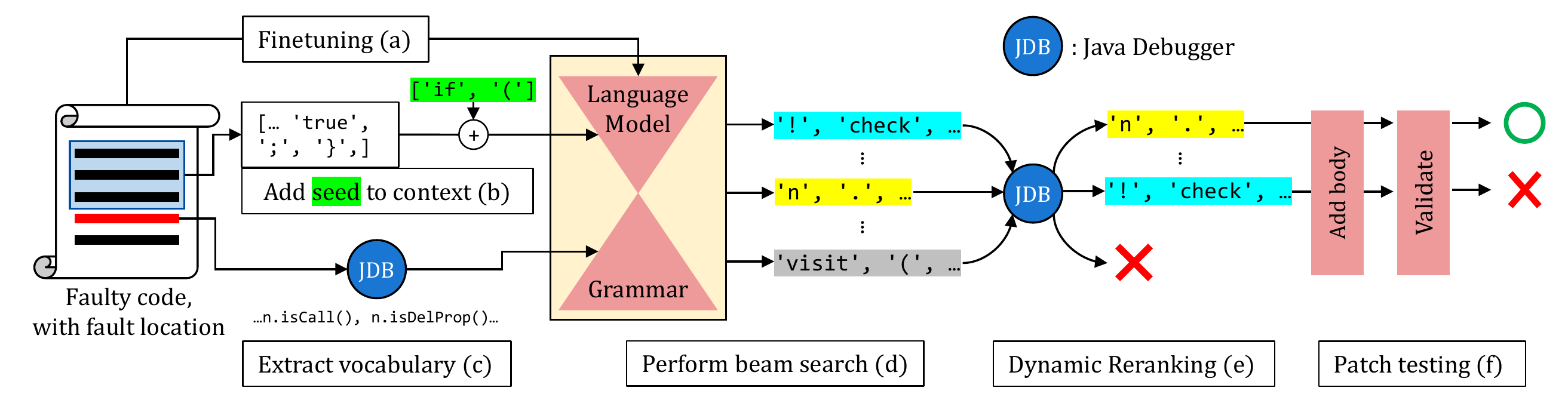}
  \caption{Diagram of \name.}
  \label{fig:overview}
\end{figure*}

\section{Approach}
\label{sec:approach}
We propose \name, a learning-based technique that specifically aims to repair \ift-statement omission faults. By moving away from the NMT paradigm (i.e., learning how to translate a faulty statement to a correct statement), and instead using the context of the faulty location to directly synthesize code, \name is well-adapted to fixing omission faults unlike existing NMT-based techniques. We explain our algorithm with the running example of the \dfj fault Closure-15, presented in Listing~\ref{lst:running_example}: the predicate \texttt{n.isDelProp()} must be synthesized and combined with the \texttt{\{return true;\}} body.

\begin{lstlisting}[language=diff, basicstyle=\ttfamily\footnotesize, caption={Closure-15 developer patch, edited for brevity.}\label{lst:running_example}]
  @@ -99,9 +99,9 @@ (...) {
1       return true;
2     }
3 +   if (n.isDelProp()) {
4 +     return true;
5 +   }
6     for (Node c = n.getFirstChild(); ...) {
7       if (...) {
\end{lstlisting}

An overview of the mechanism of \name is presented in Figure~\ref{fig:overview}. There are six steps in the repair pipeline. Before the repair proceeds, a pretrained language model is fine-tuned on the faulty version of the code so that the LM can pick up token usage patterns from the target project (Figure~\ref{fig:overview}(a)). Next, given a suspicious location, the body of the method preceding the faulty location is tokenized, and a `repair seed', namely an \ift token along with an opening parenthesis, is added to the end (Figure~\ref{fig:overview}(b)). Prior to beam search, the static information such as available identifiers and available members to types are extracted via a debugger, to construct a grammar constraining the language model (Figure~\ref{fig:overview}(c)). The fine-tuned language model will take cue from the injected \ift token and find predicate candidates using grammar-aware beam search to generate grammatically valid predicates (Figure~\ref{fig:overview}(d)). Generated predicates are evaluated by a debugger; the obtained dynamic information is used to filter and rank generated patches, based on how correct code should behave (Figure~\ref{fig:overview}(e)). Finally, we add predetermined bodies to the predicates, and evaluate the patches over tests for verification (Figure~\ref{fig:overview}(f)). The details of each step are explained in the remainder of this section.

\subsection{Preparation for Synthesis}
This subsection describes the details of finetuning, tokenization of the faulty context, and the addition of the repair seed.

\subsubsection{Finetuning}
Prior to repair, the pretrained language models are trained on the unfixed version of the project, so that the LM can `get familiar' with novel tokens (Figure~\ref{fig:overview}(a)). This allows the LM to pick up lexical patterns that are not in the pretraining dataset; for example, the token \texttt{isDelProp} from the patch of Closure-15 never appears in the pretraining dataset we use, which may hurt repair performance without finetuning. Indeed, finetuning has been shown to be useful in increasing the performance of LMs~\cite{Karampatsis2020bcbv}.

\subsubsection{Adding Repair Seed}
The finetuned LM is capable of predicting a sequence of tokens based on preceding context. To provide context, given a suspicious location we first tokenize the method content prior to the given location, resulting in a sequence of tokens representing the synthesis context. For Closure-15, Lines 1-2 are parsed to \texttt{[..., `return', `true', `;', `\}']}. Given this sequence, a `repair seed' is appended to lead the language model to generate appropriate predicates at that point (Figure~\ref{fig:overview}(b)). When the faulty location is not an \ift statement already, two tokens, \texttt{`if'} and \texttt{`('}, constitute the repair seed. As the language model likely learned that predicates appear after an \ift token is provided, given the context and the repair seed it assigns high probabilities to token sequences that resemble predicates, and thus generates predicate-resembling sequences during beam search. In the special case of when the fault location points to an \ift condition itself, we remove the last parenthesis from the \ift condition, prompting the language model to explore alternative predicates.

\subsubsection{Legal Identifier Extraction.}

A key ingredient for our grammar is the set of legal identifiers at a suspicious location. To extract them as in Figure~\ref{fig:overview}(c), we perform the following steps:

\begin{enumerate}
  \item Build the AST of the faulty file using the \texttt{javalang} library\footnote{https://github.com/c2nes/javalang} and determine the scope of each identifier based on the parent node class and the relative position of the node.
  \item Collect legal identifiers by selecting nodes such that the fixing location is located in its scope. For example, in Listing~\ref{lst:running_example} the scope of the method parameter \texttt{n} spans the entire method; thus \texttt{n} is a legal identifier.
  \item For each identifier, get additional information such as its type, bound methods, class fields, and the return type using JDB analysis. For example, we record that the parameter \texttt{n} is type \texttt{Node} and has methods such as \texttt{isDelProp()}.
  \item From within the target file, we extract static classes that were used and obtain static members of these classes.
\end{enumerate}

Additionally, we allow a predetermined set of literals. First, the primitive literals \texttt{\{null, 0, 1, true, false\}} are allowed and treated equivalently to variables that have their corresponding primitive types. These literals were chosen as they were the most common literals in our training code corpus; the use of other literals (such as large integers) is often idiosyncratic, making them inappropriate to add in our grammar rules which aim to be general. In addition to these values, we allow \name to generate a string placeholder of type \texttt{java.lang.String}; when generating concrete patches, this placeholder is replaced with strings or characters that occur within the target method.

\subsection{Beam Search}
\label{sec:beam_search}
To generate predicate candidates, we perform beam search (Figure~\ref{fig:overview}(d)), which is a commonly used algorithm that uses trained language models to synthesize a pool of likely candidates. We first describe the grammar that constrains the space that beam search can explore, then subsequently show how the grammar is integrated into beam search to effectively find likely predicates.

\subsubsection{Grammar Rules.}
To generate syntactically valid patches, we constrain the expansion of the language model: namely, it must follow grammar rules and only use the allowed vocabulary. Using the legal vocabulary set obtained in the previous section, we determine what possibilities can be explored. Specifically, to maximize the number of valid expressions generated, we keep track of the \textbf{type} of each expression, and assure that each token added is type-compatible with the preceding expression. For example, given an expression completed up to \texttt{if(n.}, we know that \texttt{n} is type \texttt{Node}, and that given the member operator (\texttt{.}) member tokens are allowed next. \texttt{Node} has members such as \texttt{isDelProp} or \texttt{next}, so we can append these tokens knowing that the ensuing expression will be valid. Using types has significant benefits, such as allowing \name to recursively keep track of which member tokens are allowed at each stage of expansion. 

\subsubsection{Grammar-directed Beam Search.} 

\begin{algorithm}[!t]
  \small
  \SetAlgoLined
  \SetKwInOut{Input}{input}\SetKwInOut{Output}{output}
  \Input{A trained language model $LM$, a token grammar $G$, context $C_{org}$, maximum length $l$ 
  and beam width $W$}
  \Output{Candidate patches \textit{CP}}
  
  $C_{init} \leftarrow$ $C_{org}$ + GetRepairSeed($C_{org}$)\;
  $CP \leftarrow$ [($C_{init}, 0$)]\;
  \For{$i$ in 1:$l$}{
    stageCandidates $\leftarrow$ []\;
    stageCandidateProbs $\leftarrow$ []\;
    \For {$C$, $p_C$ in $CP$} {
      \If {IsNotComplete($C$)} {
        newC $\leftarrow$ C\;
        legalToks $\leftarrow G(C)$\;
        \While{len(legalToks) $ = 1$} {
          newC $\leftarrow$ newC+legalToks[0]\;
          legalToks $\leftarrow G(C)$\;
        }
        legalTokProbs $\leftarrow LM(newC, legalToks)$\;
        \For {$t, p_t$ in (legalToks, legalTokProbs)} {
          stageCandidates.append($C+t$)\;
          stageCandidateProbs.append($p_C \times p_t$)\;
        }
      }
      \Else {
        stageCandidates.append($C$)\;
        stageCandidateProbs.append($p_C$)\;
      }
    }
    $CP \leftarrow$ SortBySecond(stageCandidates, stageCandidateProbs)\;
    $CP \leftarrow CP$[:W]\;
  }
  \Return $CP$\;
  \caption{Grammar-directed Beam Search}\label{alg:beam_search}
\end{algorithm}

Beam search is a heuristic algorithm that performs a pruned breadth-first search over the tree of possible sequence continuations. The pseudocode for our grammar-constrained beam search is presented in Algorithm~\ref{alg:beam_search}. The initial context to the language model is created by adding the tokenized content of the method preceding the faulty location to the repair seed (Line 1), which allows the candidate patch list ($CP$) to initialize (Line 2). The maximum number of expansions during beam search is capped by a predetermined parameter $l$ (Line 3). At each expansion step, all legal sequences stemming from the current list of candidate expressions are considered (Line 6-19). For each sequence candidate $C$ and its probability as judged by the language model $p_C$, if the current candidate is not complete (Line 7), the grammar first analyzes what tokens can follow the current expression while keeping it valid in terms of type (Line 9). 

If the grammar dictates that there is only one token possible, we append tokens to the candidate until there are multiple options (Lines 10-12). This is a novel process (not typically used in beam search) that we use in conjunction with our grammar rules to save timesteps that would be wasted expanding already determined results, thus increasing search efficiency. For example, given the expression \texttt{n.isDelProp}, we know based on type analysis that \texttt{isDelProp} accepts no arguments; Lines 10-12 allow \name to add the appropriate tokens \texttt{(} and \texttt{)} without using expansion timesteps.

Once there are multiple possible tokens to choose from, we query the LM to obtain the probability of each possible token (Line 13). For each of the legal tokens and their corresponding probabilities, we register them as potential beam candidates (Lines 14-16). At the end of each stage, we pick the top $W$ beams by descending order in probability and update the candidate patches, $CP$ (Line 20-21). The use of this grammar-aware beam search allows predicates generated by \name to almost always be syntactically correct. 

\subsection{Dynamic Reranking using JDB}
\label{sec:dyn_ranking}

To further reduce the space of patches to be evaluated, we propose a novel system based on dynamic information obtained from the Java debugger, JDB, that can eliminate and rank patches (Figure~\ref{fig:overview}(e)).

First, we observe that any correct patch \textit{must} change failing test behavior. For a side effect-free \ift statement addition patch, this means that its predicate must evaluate to \texttt{true} at some point in all failing tests. Formally, given the set of failing tests $F$ and a predicate $p$ to be added, assume a test $e \in F$ that is executed a nonzero number of times. If the set of timepoints in which $p$ is executed under test $e$ is $I_e$, and the value of $p$ at test $e$ at timepoint $i \in I_e$ is $p(e, i)$, we can denote the ``behavior must change'' criterion as in \Cref{eq:correct_nec}. Using its contrapositive, we filter out any \ift statement additions in which the predicate always evaluates to \texttt{false} in a failing test.

\begin{equation}
[p \text{ is correct}] \to \forall e \in F, \exists i \in I_e \; \text{s.t.}\; p(e, i) = T
\label{eq:correct_nec}
\end{equation}

For passing tests, there are no necessary or sufficient conditions for correct patches, due to coincidental correctness~\cite{Assi2019CoincidentalCI}. Nonetheless, a patch that minimally influences passing test behavior is intuitively more likely to be the correct patch~\cite{DeMarco2014ar}. Given the set of passing tests $P$ that cover the faulty location, the probability of an \ift statement addition with predicate $p$ being correct has an inverse correlation with PassChange$(p)$, defined in \Cref{eq:correct_prob}.

\begin{equation}
  \text{PassChange}(p) = |\{e \in P \mid \exists i \in I_e\;\text{s.t.}\;p(e, i)=T\}|
  \label{eq:correct_prob}
\end{equation}

Based on these two principles, we derive the score for the ordering of predicates to use in \ift-statement insertion in \Cref{eq:pred_score}. As the principle is based on behavior \textit{change}, how likely a predicate is correct depends on how the predicate is used as well. If the predicate is part of an \ift-statement complete with a new body added, the predicate must evaluate to \texttt{true} in failing tests, while it is good if it evaluates to \texttt{false} in passing tests. On the other hand, if the predicate is used to wrap existing code, the opposite holds. As such, we give multiple scores to each predicate, each corresponding to a body template described in \Cref{sec:patch_validation}. In practice, \Cref{eq:pred_score} can lead to many ties; these are broken by the language model's likelihood score for each predicate.

\begin{equation}
  \text{Score}(p) = \begin{cases}
    - \infty & \text{if}\: \exists e \in F, \forall i \in I_e \; p(e, i) = F \\
    - \text{PassChange}(p) & \text{else}
  \end{cases}
  \label{eq:pred_score}
\end{equation}

To efficiently calculate predicate scores using \Cref{eq:pred_score}, \name again uses the \textbf{debugger}. The debugger stops execution at the fix location, allowing evaluation of generated expressions. Our debugger use consists of two steps: one short pass to check which expressions generated by the LM are actually Boolean and have no side effects (the \textit{validation pass}), and one longer pass to evaluate predicate score according to \Cref{eq:pred_score} (the \textit{ranking pass}). As there is no compile overhead, and the context preceding the faulty location does not need to be executed multiple times, using the debugger in the proposed manner allows a significant speedup relative to executing each patch individually. The empirical effects of this ranking are investigated in Section~\ref{sec:rq_ablation}.

\subsection{Patch Validation}
\label{sec:patch_validation}
At this point, we have generated and ranked a set of predicates; we must now determine the bodies for these predicates to complete a patch and validate it based on the test suite (Figure~\ref{fig:overview}(f)).
\paragraph{Attaching a Body}
We apply three templates for bodies: (BT1) returning a literal or raising an exception within the failing test (as does ACS~\cite{Xiong2017acs}), (BT2) wrapping the code that follows the faulty location and is within the same block with the predicate (as in TBar~\cite{Liu2019tbar} FP4.4), and (BT3) adding a clause to an existing predicate, in which case there is no need to generate a new body. In Closure-15, using BT1 allows making the correct patch. As our focus is on predicate synthesis, we do not use complex techniques to synthesize the body; however, the language model is theoretically capable of synthesizing statements as well, and as such synthesizing the body could be explored in future work.
\paragraph{Validation}
All generated patches are sorted in order of descending score and evaluated relative to the test suite provided by \dfj. A patch that passes all tests is \textit{plausible}; plausible patches are manually inspected to check for correctness. If the patch is judged to be semantically equivalent to the developer patch, the patch is deemed \textit{correct}.

\section{Experimental Setup}
\label{sec:expr_details}

\subsection{Configurations}

\subsubsection{Fault Localization} We experiment with two FL settings: perfect statement fault localization and method-given fault localization. Perfect fault localization allows measuring repair performance without localization bias~\cite{Liu2020ot}; further it is widely used in prior learning-based APR techniques~\cite{Lutellier2020cc,Lutellier2021ca,Chen2019SequenceRSL,Chakraborty2020CODIT}, allowing for fair comparison. Method-given fault localization is also used to demonstrate the versatility of \name, and for fair comparison with state-of-the-art constraint-based techniques~\cite{Chen2018JAID,Xu2020Restore}. In this formulation, the buggy method is known, and \name iterates over fix locations within the method using the Ochiai~\cite{Abreu2007Ochiai} suspiciousness order.

\subsubsection{Language Model} The LM is pretrained on the Java-med dataset gathered by Alon et al.~\cite{alon2018codeseq}. We extract methods using \texttt{javalang} and treat them as sentences to pretrain the LM; we discard the less than 1\% of methods and files that \texttt{javalang} fails to parse. We perform tokenization using \texttt{javalang}, and further perform subtokenization using Byte-Pair Encoding (BPE)~\cite{Karampatsis2020bcbv} with approximately 5,000 pairs, so that the LM does not suffer from out-of-vocabulary issues. Prior to training, data related to the \dfj projects is purged. The LM is trained for five epochs on the subtokenized pretraining dataset, which is enough for learning to stabilize. Pretraining is an offline process that does not need to be repeated for each fault. We use GRUs ~\cite{Cho2014LearningPR}, as they are known to be better than traditional recurrent neural networks and comparable in performance to LSTMs~\cite{Chung2014EmpiricalEO} while being structurally simple. The detailed architecture of our model is provided in the Appendix.

\subsubsection{GLAD hyperparameters} During finetuning (\Cref{fig:overview} (a)), the language model is trained for one epoch on the faulty code, as more training can quickly lead to overfitting due to the smaller size of the faulty project. We perform beam search with a beam width $W$ of 10,000 and maximum length $l$ of 15 tokens, which empirically leads to the best performance. The beam width is larger than prior work~\cite{Lutellier2021ca}; \name can navigate this larger space thanks to dynamic reranking. Dynamic reranking (\Cref{fig:overview} (e)) is given a timeout of 15 minutes, which was empirically determined to be sufficient. We run \name for a maximum of three hours, although \name often completes much earlier than that. After patch verification the plausible patches are inspected for correctness, as is done in prior work~\cite{Chen2018JAID,Lutellier2021ca}. We evaluate on 95 bugs from \dfj v1.2 and 89 bugs newly introduced in \dfj v2.0 for a total of 184 bugs that include \ift statement faults.

\subsubsection{Computational Resources} Experiments were run on eight-core machines with Intel(R) Core(TM) i7-6700 CPU and NVIDIA 1080/3090 GPUs. 

\subsection{Research Questions}
\label{sec:all_rqs}

\noindent\textbf{RQ1. Repair Effectiveness:} How many omission faults can \name successfully repair, and how distinct is its repair profile? Using the two fault localization scenarios, we compare the performance of \name on the widely studied \dfj v1.2 with multiple baselines. To emphasize how unique the suite of bugs that \name fixes are, we first compare our performance using perfect fault localization with (i) TBar employing perfect-FL~\cite{Liu2019tbar}, which represents an ensemble of multiple template-based tools, and (ii) \textit{all} reported fixes of deep learning-based tools \cite{Lutellier2021ca, Lutellier2020cc, Chen2019SequenceRSL,Li2020DLFixCB,Hata2018Ratchet, Chakraborty2020CODIT, Zhu2021Recoder} that we are aware of. We additionally compare our method-given repair performance with the state-of-the-art condition generating repair tool Restore~\cite{Xu2020Restore}, which evaluates under equivalent settings. Finally, we report the number of times each bug was fixed by a technique listed in \Cref{tab:all_tools} and check how many bugs \name has fixed against the field as a whole. If \name can fix previously unfixable bugs, it would indicate that \name is indeed unique in its repair capabilities. 

\noindent\textbf{RQ2. Generalization:} In order to mitigate external validity 
concerns and show that \name can operate over a wide range of projects, we 
evaluate \name using perfect fault localization on the buggy projects that were 
newly introduced in \dfj v2.0. By verifying its performance against bugs in new 
projects, we hope to show that \name is a general-purpose technique for 
handling \ift-omission faults that is not overfitted to the widely studied \dfj 
v1.2 benchmark.

\noindent\textbf{RQ3. Ablation Study:} How much contribution does each of the components of \name make to repair? To investigate, we remove each component of \name and compare the performance to the full model, using all 184 bugs studied in RQ1 and RQ2, as follows:
\begin{itemize}
\item\textbf{-Finetuning.} Finetuning can simply be omitted from \name, as it is not intertwined with other repair components.
\item\textbf{-Grammar.} Grammar can be removed by changing $G$ in Line 9 of Algorithm~\ref{alg:beam_search} to allow all tokens, resulting in unrestricted beam search based on token naturalness alone.
\item\textbf{-Language Model.} We modify Line 20 of Algorithm~\ref{alg:beam_search} to sort at random, instead of using the LM.
\end{itemize}

\noindent Additionally, we analyze how much verification effort \textbf{dynamic ranking} is saving, based on how the correct patch's ranking improved relative to when only using the likelihood score provided by the LM.

\noindent\textbf{RQ4. Efficiency:} How efficient is \name in comparison to other APR techniques? To investigate, we gather statistics about the runtime of \name, and compare them with two other comparable techniques that report their average runtime, CURE~\cite{Lutellier2021ca} and Restore~\cite{Xu2020Restore}.

\noindent\textbf{RQ5. Qualitative Analysis:} What can we learn from both the successful and unsuccessful patches generated by \name? With RQ5, we qualitatively analyze how \name generates successful patches. Further, we identify failure instances of \name to highlight shortcomings and potential improvements in future work.

\section{Results}
\label{sec:results}

\subsection{Repair Performance (RQ1)}

\begin{table}[ht]
  \centering
  \caption{Faults correctly fixed by \name in \dfj v1.2. $\bigcirc$/$\triangle$/$\times$ stand for correctly/plausibly/not fixed, respectively. \label{tab:all_dom_bugs} The +C column shows the number of characters added by the developer patch for each bug, and the DL column represents the known fix results of learning-based APR techniques. On the right, GLAD-M shows the performance of GLAD when the method is given, while RSTR is short for Restore. Note the DL column does not have plausible results, as many tools do not report their per-bug plausible fixes.}
  \scalebox{0.95}{
  \begin{tabular}{ccccc|ccc}
  \toprule

  Fault ID & +C & \name & TBar & DL & \name-M & RSTR & Ever \\\midrule
  C-4 & 17 & $\bigcirc$ & $\bigcirc$ & $\times$ & $\triangle$ & $\times$ & 9 \\
  C-9 & 37 & $\bigcirc$ & $\bigcirc$ & $\bigcirc$ & $\triangle$ & $\times$ & 11 \\
  C-26 & 21 & $\bigcirc$ & $\bigcirc$ & $\bigcirc$ & $\bigcirc$ & $\bigcirc$ & 15 \\
  Cl-1 & 29 & $\bigcirc$ & $\times$ & $\times$ & $\bigcirc$ & $\times$ & 0 \\
  Cl-5 & 39 & $\bigcirc$ & $\times$ & $\times$ & $\bigcirc$ & $\bigcirc$ & 1 \\
  Cl-15 & 33 & $\bigcirc$ & $\times$ & $\times$ & $\bigcirc$ & $\times$ & 0 \\
  Cl-18 & 45 & $\bigcirc$ & $\bigcirc$ & $\bigcirc$ & $\times$ & $\bigcirc$ & 9 \\
  Cl-33 & 33 & $\bigcirc$ & $\times$ & $\bigcirc$ & $\bigcirc$ & $\bigcirc$ & 4 \\
  Cl-38 & 19 & $\bigcirc$ & $\bigcirc$ & $\bigcirc$ & $\bigcirc$ & $\times$ & 9 \\
  Cl-52 & 24 & $\bigcirc$ & $\times$ & $\times$ & $\bigcirc$ & $\times$ & 0 \\
  Cl-57 & 41 & $\bigcirc$ & $\times$ & $\bigcirc$ & $\triangle$ & $\times$ & 4 \\
  Cl-104 & 20 & $\bigcirc$ & $\times$ & $\bigcirc$ & $\bigcirc$ & $\times$ & 2 \\
  Cl-113 & 29 & $\bigcirc$ & $\times$ & $\times$ & $\times$ & $\bigcirc$ & 1 \\
  Cl-118 & 39 & $\bigcirc$ & $\times$ & $\bigcirc$ & $\bigcirc$ & $\bigcirc$ & 3 \\
  Cl-125 & 33 & $\bigcirc$ & $\triangle$ & $\times$ & $\bigcirc$ & $\bigcirc$ & 1 \\
  Cl-128 & 36 & $\bigcirc$ & $\times$ & $\times$ & $\bigcirc$ & $\bigcirc$ & 1 \\
  Cl-130 & 35 & $\bigcirc$ & $\times$ & $\times$ & $\bigcirc$ & $\bigcirc$ & 1 \\
  L-9 & 185 & $\times$ & $\times$ & $\times$ & $\bigcirc$ & $\times$ & 0 \\
  L-24 & 24 & $\bigcirc$ & $\bigcirc$ & $\times$ & $\times$ & $\times$ & 8 \\
  L-33 & 25 & $\bigcirc$ & $\bigcirc$ & $\bigcirc$ & $\bigcirc$ & $\bigcirc$ & 14 \\
  L-39 & 68 & $\bigcirc$ & $\bigcirc$ & $\times$ & $\times$ & $\times$ & 4 \\
  L-45 & 49 & $\times$ & $\times$ & $\times$ & $\bigcirc$ & $\times$ & 1 \\
  M-25 & 86 & $\bigcirc$ & $\times$ & $\times$ & $\bigcirc$ & $\times$ & 2 \\
  M-28 & 95 & $\bigcirc$ & $\triangle$ & $\times$ & $\triangle$ & $\times$ & 0 \\
  M-48 & 48 & $\bigcirc$ & $\times$ & $\times$ & $\triangle$ & $\times$ & 0 \\
  M-94 & 15 & $\bigcirc$ & $\triangle$ & $\bigcirc$ & $\times$ & $\times$ & 5 \\
  Mo-8 & 40 & $\bigcirc$ & $\times$ & $\bigcirc$ & $\times$ & $\times$ & 2 \\
  Mo-22 & 34 & $\times$ & $\times$ & $\times$ & $\bigcirc$ & $\times$ & 1 \\
  Mo-24 & 58 & $\bigcirc$ & $\times$ & $\times$ & $\bigcirc$ & $\times$ & 0 \\
  Mo-34 & 49 & $\bigcirc$ & $\times$ & $\times$ & $\bigcirc$ & $\times$ & 0 \\
  Mo-38 & 75 & $\bigcirc$ & $\bigcirc$ & $\bigcirc$ & $\times$ & $\times$ & 10 \\

  \midrule
  Total & - & 28 & 9 & 12 & 19 & 10 & - \\
  \bottomrule
  \end{tabular}
  }
\end{table} 

Repair results on faults from \dfj v1.2 are presented in \Cref{tab:all_dom_bugs}. \name can fix 28 faults correctly using perfect FL, and 19 faults with method-given FL. More to the point, however, is how \textit{different} the bugs that \name fixes are. Relative to TBar, \name fixes \textbf{19} different bugs correctly, confirming that \name fixes a large group of bugs with which template-based techniques struggle. Further, when comparing with the \textit{entire} set of bugs known to be fixed by deep learning-based techniques (DL), \name fixes \textbf{16} different bugs under the perfect fault localization scenario, significantly expanding what is possible with learning-based APR techniques. While other techniques do fix bugs that \name fixes in certain instances, they generally fix bugs that add expressions to an already existing statement (weak omission faults), or common cases such as null checks. This trend continues when \name is evaluated without the exact fix location (GLAD-M) under the method-given setting against Restore (RSTR); \name fixes 11 bugs correctly that Restore could not under equivalent settings, suggesting that \name is indeed capable of fixing a unique set of bugs. Even when compared with all 40 APR tools we are aware of (Ever), overall \name fixes \textbf{eight} bugs for the first time, underscoring the uniqueness of \name-generated solutions. This is because these bugs often require complex expressions to be synthesized (as is qualitatively explored in RQ5), which are uniquely handled by \name. 
Finally, note the bugs that \name fixes are generally within the 16-64 added character range (`next level' bugs in \Cref{sec:apr_difficulty}), suggesting \name met its stated goal.

\begin{tcolorbox}[boxsep=2pt,left=2pt,right=2pt,top=1pt,bottom=1pt]
  \textbf{Answer to RQ1:} \name can fix a unique set of bugs, regardless of the comparison group, fixing 16 bugs that previous learning-based APR techniques could not fix and 8 that were never reported to be correctly fixed. Further, the bugs that \name fixes require some degree of synthesis, suggesting \name met its original mission.
\end{tcolorbox}

\begin{table}[ht]
  \centering
  \caption{Number of faults fixed by \name in \dfj v2.0. Baseline numbers marked by an asterisk are from Zhu et al.~\cite{Zhu2021Recoder}, and are not directly comparable to those of \name. \label{tab:d4jv2_rep}}
  \scalebox{1.0}{
  \begin{tabular}{{r}*{4}{c}}
  \toprule
   Patch type & TBar* & SimFix* & Recoder* & \name \\\midrule
   Plausible  & 50 & 25 & 46 & 34 \\
   Correct    & 8 & 2 & 19 & 20 \\
  \bottomrule
  \end{tabular}
  }
\end{table} 

\subsection{Generalization (RQ2)}

The number of bugs that \name plausibly and correctly fixed within the \dfj v2.0 dataset are presented in \Cref{tab:d4jv2_rep}. As shown, \name continues to fix a significant number of bugs, even though previous work suggests bugs added in \dfj v2.0 are more difficult to repair~\cite{Zhu2021Recoder}. While \name is shown to correctly fix the largest number of bugs in the table above, these numbers are not directly comparable as Zhu et al.~\cite{Zhu2021Recoder} used GZoltar-based FL for APR evaluation, which were not replicated in this study for consistency with RQ1 and due to the fact that GZoltar failed on certain \dfj v2.0 bugs. Nonetheless, it is noteworthy that, while Zhu et al.~\cite{Zhu2021Recoder} report the performance of TBar and SimFix dropping significantly when going from \dfj v1.2 to v2.0, the performance of \name is roughly consistent. As such, we would like to emphasize that \name continues to successfully generate patches with complex expressions in different projects: for example, \name generates the correct predicate \texttt{(l<Integer.MIN\_VALUE||l>Integer.MAX\_VALUE)} for the bug Compress-46, despite the fact that the exact predicate form never appears within the project.

\begin{tcolorbox}[boxsep=2pt,left=2pt,right=2pt,top=1pt,bottom=1pt]
  \textbf{Answer to RQ2:} \name successfully generalizes to new bugs introduced in \dfj v2.0, showing its versatility and generality as a repair tool.
\end{tcolorbox}

\subsection{Component Ablation Study (RQ3)}
\label{sec:rq_ablation}

The performance of each ablated model is presented in \Cref{tab:ablation}. The Plaus. and Corr. columns represent the number of faults fixed plausibly and correctly. The time presented on the Time column corresponds to the median time (i) finetuning the model on the target project (\Cref{fig:overview}(a)), (ii) extracting the legal tokens and constructing the grammar (\Cref{fig:overview}(c)), and (iii) performing beam search (\Cref{fig:overview}(d)).

\begin{table}[ht]
  \centering
  \caption{Faults fixed by \name with ablated models. \label{tab:ablation}}
  \scalebox{0.9}{
  \begin{tabular}{{r}*{3}{c}}
  \toprule
   Method & Plaus. & Corr. & Time\\\midrule
   \name & 75 & 48 & - \\
  -Finetuning   & 71 & 39 & 34s \\
  -Grammar & 48 & 21 & 4s \\
  -Language Model & 13 & 1 & 69s \\
  \bottomrule
  \end{tabular}
  }
\end{table} 

We find that all components that directly affect how the patches are made contribute to repair performance. First, without finetuning, the model generates a similar number of plausible patches, but the total number of faults correctly fixed drops by nine. These bugs generally involve tokens that are not within the pretraining dataset, indicating that finetuning indeed helps the language model pick up within-project lexical patterns. It is worth noting, however, that the performance drop is less severe than the other ablated models; thus when using a larger LM that is too expensive to retrain, it may be acceptable to use it without finetuning. Without grammar, the number of correctly fixed faults drops by about half, while the number of plausible patches drops by about a third. As the LM is unaware of what expressions are acceptable within the given project, its capabilities drop significantly, often failing to generate necessary but uncommon tokens. This suggests that LMs, and perhaps learning-based techniques in general, would greatly benefit from project-specific static information that can easily be acquired: incorporating such information may improve the performance drastically at a low computational cost. Without the LM (essentially random sampling from grammar-conforming expressions), the number of plausibly fixed faults drops by 83\% to only 13 faults, while only one bug is correctly fixed. Thus, while our grammar is powerful enough to fix some bugs with weak test suites, the grammar alone does not lead to strong performance, and guidance from the LM is necessary. Overall, these results show that the synergy between the LM and the grammar enables \name to generate complex expressions that fix previously unfixable faults.

\begin{table}[ht]
  \centering
  \caption{Ranking performance of \name regarding the correct patch with and without dynamic reranking (DR). MRR is Mean Reciprocal Rank, acc@n denotes the correct patches found within the top $n$ patches, Max. shows the worst ranking of the correct patch over all faults correctly fixed, and Time shows the execution time of the validity pass and ranking pass separately.\label{tab:dr_ablation}}
  \scalebox{0.9}{
  \begin{tabular}{{r}*{5}{c}}
  \toprule
        & MRR & acc@1 & acc@5 & Max. & Time \\\midrule
  \name & 0.449 & 18 & 25 & 745 & 13.5+119s \\
  -DR & 0.138 & 5 & 10 & 28321 & - \\
  \bottomrule
  \end{tabular}
  }
\end{table}

Meanwhile, Table~\ref{tab:dr_ablation} shows how our dynamic ranking enhances the efficiency of \name. The verification effort needed to find the correct expression decreases significantly; acc@1 and acc@5 increase from 5 and 10 to 18 and 25, respectively. Importantly, thanks to the dynamic ranking, the worst cases get substantially more manageable. In the most drastic example, the correct patch would have been placed at 28,321th place to be validated, but dynamic ranking moves it up to the third place, reducing the necessary effort by 4 degrees of magnitude and saving tens of hours in verification time. This improvement, which allows \name to be time-efficient as demonstrated in RQ4, is obtained by investing a little over two minutes: 13.5 seconds on the quick validity pass, and two minutes on dynamic ranking. 
Overall, dynamic ranking also plays an important role in making the repair cost of \name feasible even as it generates unique expressions.

\begin{tcolorbox}[boxsep=2pt,left=2pt,right=2pt,top=1pt,bottom=1pt]
  \textbf{Answer to RQ3:} Each component of \name is essential, contributing independently to the final performance of the model in a manner that is expected by their roles.
\end{tcolorbox}

\begin{table}[ht]
  \centering
  \caption{Statistics regarding time (in minutes) until the first plausible patch is generated using \name.\label{tab:time_stats}}
  \scalebox{0.9}{
  \begin{tabular}{{r}*{3}{c}}
  \toprule
   Repair type         & Min. & Med. & Mean \\\midrule
   Perfect FL + \name   & 1.72 & 9.38 & 17.76 \\
   Perfect FL + CURE~\cite{Lutellier2021ca} & >2.5 & - & 16.5 \\
   Method-given + \name & 3.08 & 18.59 & 38.65 \\
   Method-given + Restore~\cite{Xu2020Restore} & 1.1 & 10.4 & 38.25 \\
  \bottomrule
  \end{tabular}
  }
\end{table} 

\subsection{Runtime (RQ4)}

While the timeout given to \name is three hours per bug, we find that the actual time taken to generate and verify a patch is much shorter in general. The runtime required to fix a bug, along with comparable baselines, is provided in \Cref{tab:time_stats}. (The minimum fix time of CURE is the amount of time they state it takes to \textit{generate} patches using beam search.) Looking at the Mean column, we find that \name is capable of generating plausible patches in roughly the same amount of time when compared to baselines under both FL settings; indeed, even without the exact fault location, \name can generate patches for more than half the bugs it plausibly fixed in less than 20 minutes. The bugs fixed in this short time include relatively complex predicates, as is shown in RQ5. These results suggest that (i) \name is indeed effective at fixing complex bugs, and that (ii) \name would be effective as part of an APR ensemble to run when other APR tools fail.

\begin{tcolorbox}[boxsep=2pt,left=2pt,right=2pt,top=1pt,bottom=1pt]
  \textbf{Answer to RQ4:} \name generates most of its correct patches in a short time, suggesting \name is efficient when repairing bugs.
\end{tcolorbox}

\subsection{Qualitative Analysis (RQ5)}
We look at examples of successful and unsuccessful patches generated by \name to answer RQ5.

\begin{table}[ht]
  \footnotesize
  \centering
  \caption{Correct patches for bugs only fixed by \name. \texttt{-dev} represents the developer patch for the corresponding bug. \label{tab:succ_example}}
  \scalebox{1.0}{
  \begin{tabular}{p{0.17\linewidth}>{\raggedright\arraybackslash}p{0.70\linewidth}}
  \toprule
  Patch ID & Patch Content \\
  \midrule
  Cl-52-dev & Append \texttt{\&\&s.charAt(0)!=`0'} to \texttt{return len>0}\\
  \rowcolor{green} Cl-52-\name & Add \texttt{if(len>0\&\&s.charAt(0)==`0') \{return false;\}}\\\hline
  % L-9-dev & Add \texttt{if(patternMatcher.regionStart() != patternMatcher.regionEnd()) \{throw new IllegalArgumentException("...");\}} \\
  % \rowcolor{green} L-9-\name & Add \texttt{if(patternMatcher.regionStart() != patternMatcher.regionEnd()) \{return;\}} \\\hline
  Mo-24-dev & Add \texttt{if(invocation.getMock() == invocation.getArguments()[0]) \{return 0;\}}\\
  \rowcolor{green} Mo-24-\name & Add \texttt{if(invocation.getMock() == invocation.getArguments()[0]) \{return 0;\}}\\ \hline
  Mo-34-dev & Append \texttt{\&\&i.getArguments().length>k} to condition \\
  \rowcolor{green} Mo-34-\name & Add \texttt{if(k==i.getArgumentsCount()) \{break;\}} \\
  \bottomrule
  \end{tabular}
  }
\end{table}

\textbf{Successful Cases.} Correct patches that only \name can generate are presented in \Cref{tab:succ_example}. All such bugs require the synthesis of a relatively long predicate, which makes them fundamentally challenging for existing template-based and learning-based techniques. Due to the large number of operators and identifiers/constants required to generate correct expressions, they pose a challenge to constraint-based techniques as well, as a consequence of combinatorial space explosion. 
Regarding each fault, Closure-52 compares the value of a method call with an argument to a character; to the best of our knowledge, no learning-based technique has ever synthesized patches requiring characters or string literals. 
For Mockito-24, the patch requires a complex expression that involves the comparison of the results of two distinct method calls. Further, the return type of \texttt{invocation.getArguments()} is an array, and thus the patch needs to access the array element to make the correct predicate. The fact that \name can successfully generate this expression is a testament to how grammar can help the language model during synthesis. 
Finally, in Mockito-34, \name finds the method \texttt{getArgumentsCount()}, which is more concise than the developer-written expression \texttt{getArguments().length} while being semantically equivalent; indeed, the LM prioritized using the available method. Again, due to the relatively large number of distinct identifiers involved in the expression (\texttt{i}, \texttt{k}, \texttt{getArgumentsCount}), which also need to be appropriately matched using correct operators to generate the correct expression, such a bug requires precise expression synthesis to be fixed.

As far as we know, these bugs are challenging for existing APR tools to fix even with a large computational budget, due to their complexity and components; yet, these complex bugs are fixed by \name in less than thirty minutes. As such, these results demonstrate that \name makes a significant contribution to the state-of-the-art in APR.

\begin{table}[ht]
  \footnotesize
  \centering
  \caption{Incorrect patches  generated by \name. \label{tab:fail_example}}
  \scalebox{1.0}{
  \begin{tabular}{p{0.1\linewidth}>{\raggedright\arraybackslash}p{0.8\linewidth}}
  \toprule
  Patch ID & Patch Content \\
  \midrule
  M-106-dev & Add \texttt{if(num.intValue() < 0) \{ pos.setIndex(initialIndex); return null; \}}\\
  \rowcolor{pink} M-106-\name & Add \texttt{if(num.intValue() < 0) \{ return null; \}}\\\hline
  % L-9-dev & Add \texttt{if(patternMatcher.regionStart() != patternMatcher.regionEnd()) \{throw new IllegalArgumentException("...");\}} \\
  % \rowcolor{green} L-9-\name & Add \texttt{if(patternMatcher.regionStart() != patternMatcher.regionEnd()) \{return;\}} \\\hline
  % Mo-24-dev & Add \texttt{if(invocation.getMock() == invocation.getArguments()[0]) \{return 0;\}}\\
  % \rowcolor{green} Mo-24-\name & Add \texttt{if(invocation.getMock() == invocation.getArguments()[0]) \{return 0;\}}\\ \hline
  T-27-dev & Wrap block with \texttt{if(sep.iAfterParser==null \&\& sep.iAfterPrinter==null)} \\
  \rowcolor{pink} T-27-\name & Add \texttt{if(size <= sep.getClass().getInterfaces().length ) \{ return null; \}} \\
  \bottomrule
  \end{tabular}
  }
\end{table}

\textbf{Failing Cases.} We present informative cases in which \name fails to generate the correct patch, shown in \Cref{tab:fail_example}. In Math-106, we find that \name has actually generated the correct predicate and return statement. Unfortunately, the patch is not correct, as the statement \texttt{pos.setIndex(initialIndex);} was not added to the body of the \ift statement. This is in fact a rather common situation: we find that \name generates the correct predicate for 32 bugs that it did not fix correctly. Thus along with the 48 bugs it fixes correctly, \name generates the correct predicate for 80 bugs. This suggests that an even greater number of omission faults could be solved by further research into body synthesis; we hope to investigate this prospect in follow-up research.

Regarding Time-27, note that the predicate requires synthesizing specific field accesses, such as \texttt{`sep.iAfterPrinter'}. This is difficult for the language model, as (i) the token \texttt{`iAfterPrinter'} never appears in the LM training data, and (ii) the token is never accessed as a field within the Time project except in this instance. Thus the LM assigns a low probability to the \texttt{`sep.iAfterPrinter'} sequence, causing beam search to drop the candidate. Note that this is a weakness that many other techniques share. For example,  techniques that use code snippets from other parts of the project~\cite{Goues2012tse,Xu2020Restore} would equally struggle here due to the lack of precedents.

\begin{tcolorbox}[boxsep=2pt,left=2pt,right=2pt,top=1pt,bottom=1pt]
  \textbf{Answer to RQ5:} A qualitative inspection reveals \name is capable of synthesizing expressions that other techniques would have a great deal of difficulty doing so. Meanwhile, \name struggles when there is little lexical precedent for a predicate, as many other repair tools do.
\end{tcolorbox}

\section{Threats to Validity}
\label{sec:ttv}
\textbf{Internal validity} relates to whether the experiments performed in the study have precluded other possible explanations. In RQ1, we compare with a large body of existing APR techniques to find that \name can fix multiple faults that were not fixed by the community despite multiple attempts by 40 different tools, suggesting that \name is capable of fixing omission faults that pose a challenge to other techniques. In RQ3, we establish that each component has a significant effect on the final performance, minimizing the effects from randomness.

\noindent \textbf{External validity} relates to whether the results of this work will generalize to different subjects. We find that \name can fix patches over multiple repositories written in Java: its performance was consistent under similar settings when applied to different projects, as demonstrated in RQ1/RQ2. Such results suggest \name can successfully repair over a wide range of projects. Nonetheless, further research is necessary to understand how far the performance of \name generalizes. While the principles of \name are not derived from any characteristics of the Java language itself, and as such it is possible to implement \name on other programming languages, the performance of \name under such conditions is currently unknown.

\section{Conclusion}
\label{sec:conclusion}
In this work, we analyze existing APR literature to identify which faults APR tools should tackle next. Our results indicate that many of the unfixed faults involve omission faults, and a sizable portion of those omission faults are \ift-statement omission faults. With this in mind, we propose \name, an APR technique that, given a fault location, uses language models, a debugger, and grammar-based beam search to synthesize and rank natural predicate candidates. We find that \name is highly orthogonal to existing work, fixing 16 faults never fixed by previous learning-based tools, and fixing eight faults no other tool has fixed. The performance is consistent even when evaluated over a diverse set of projects, underscoring the versatility of \name, while fixing bugs often takes less than thirty minutes. We thus verify the utility of \name as a repair tool that is effective at fixing bugs within its domain; the orthogonality of \name makes it a good candidate for an ensemble of APR tools to deploy as well. Our results invite further investigations into both the capability of language models, and how to harness them to achieve strong repair performance.

\bibliographystyle{ACM-Reference-Format}
\bibliography{acmart}


\begin{thebibliography}{62}

%%% the Web address.

\ifx \showCODEN    \undefined \def \showCODEN     #1{\unskip}     \fi
\ifx \showDOI      \undefined \def \showDOI       #1{#1}\fi
\ifx \showISBNx    \undefined \def \showISBNx     #1{\unskip}     \fi
\ifx \showISBNxiii \undefined \def \showISBNxiii  #1{\unskip}     \fi
\ifx \showISSN     \undefined \def \showISSN      #1{\unskip}     \fi
\ifx \showLCCN     \undefined \def \showLCCN      #1{\unskip}     \fi
\ifx \shownote     \undefined \def \shownote      #1{#1}          \fi
\ifx \showarticletitle \undefined \def \showarticletitle #1{#1}   \fi
\ifx \showURL      \undefined \def \showURL       {\relax}        \fi
% invisible to TeX
\providecommand\bibfield[2]{#2}
\providecommand\bibinfo[2]{#2}
\providecommand\natexlab[1]{#1}
\providecommand\showeprint[2][]{arXiv:#2}

\bibitem[\protect\citeauthoryear{Abreu, Zoeteweij, and van Gemund}{Abreu
  et~al\mbox{.}}{2007}]%
        {Abreu2007Ochiai}
\bibfield{author}{\bibinfo{person}{Rui Abreu}, \bibinfo{person}{Peter
  Zoeteweij}, {and} \bibinfo{person}{Arjan~J.C. van Gemund}.}
  \bibinfo{year}{2007}\natexlab{}.
\newblock \showarticletitle{On the Accuracy of Spectrum-based Fault
  Localization}. In \bibinfo{booktitle}{\emph{Testing: Academic and Industrial
  Conference Practice and Research Techniques - MUTATION (TAICPART-MUTATION
  2007)}}. \bibinfo{pages}{89--98}.
\newblock
\urldef\tempurl%
\url{https://doi.org/10.1109/TAIC.PART.2007.13}
\showDOI{\tempurl}


\bibitem[\protect\citeauthoryear{Alon, Brody, Levy, and Yahav}{Alon
  et~al\mbox{.}}{2019}]%
        {alon2018codeseq}
\bibfield{author}{\bibinfo{person}{Uri Alon}, \bibinfo{person}{Shaked Brody},
  \bibinfo{person}{Omer Levy}, {and} \bibinfo{person}{Eran Yahav}.}
  \bibinfo{year}{2019}\natexlab{}.
\newblock \showarticletitle{code2seq: Generating Sequences from Structured
  Representations of Code}. In \bibinfo{booktitle}{\emph{International
  Conference on Learning Representations}}.
\newblock
\urldef\tempurl%
\url{https://openreview.net/forum?id=H1gKYo09tX}
\showURL{%
\tempurl}


\bibitem[\protect\citeauthoryear{Assi, Trad, Maalouf, and Masri}{Assi
  et~al\mbox{.}}{2019}]%
        {Assi2019CoincidentalCI}
\bibfield{author}{\bibinfo{person}{R.~A. Assi}, \bibinfo{person}{Chadi Trad},
  \bibinfo{person}{Marwan Maalouf}, {and} \bibinfo{person}{W. Masri}.}
  \bibinfo{year}{2019}\natexlab{}.
\newblock \showarticletitle{Coincidental correctness in the Defects4J
  benchmark}.
\newblock \bibinfo{journal}{\emph{Software Testing}}  \bibinfo{volume}{29}
  (\bibinfo{year}{2019}), \bibinfo{pages}{n/a}.
\newblock


\bibitem[\protect\citeauthoryear{Brown, Mann, Ryder, Subbiah, Kaplan, Dhariwal,
  Neelakantan, Shyam, Sastry, Askell, Agarwal, Herbert-Voss, Krueger, Henighan,
  Child, Ramesh, Ziegler, Wu, Winter, Hesse, Chen, Sigler, Litwin, Gray, Chess,
  Clark, Berner, McCandlish, Radford, Sutskever, and Amodei}{Brown
  et~al\mbox{.}}{2020}]%
        {Brown2020GPT2}
\bibfield{author}{\bibinfo{person}{Tom Brown}, \bibinfo{person}{Benjamin Mann},
  \bibinfo{person}{Nick Ryder}, \bibinfo{person}{Melanie Subbiah},
  \bibinfo{person}{Jared~D Kaplan}, \bibinfo{person}{Prafulla Dhariwal},
  \bibinfo{person}{Arvind Neelakantan}, \bibinfo{person}{Pranav Shyam},
  \bibinfo{person}{Girish Sastry}, \bibinfo{person}{Amanda Askell},
  \bibinfo{person}{Sandhini Agarwal}, \bibinfo{person}{Ariel Herbert-Voss},
  \bibinfo{person}{Gretchen Krueger}, \bibinfo{person}{Tom Henighan},
  \bibinfo{person}{Rewon Child}, \bibinfo{person}{Aditya Ramesh},
  \bibinfo{person}{Daniel Ziegler}, \bibinfo{person}{Jeffrey Wu},
  \bibinfo{person}{Clemens Winter}, \bibinfo{person}{Chris Hesse},
  \bibinfo{person}{Mark Chen}, \bibinfo{person}{Eric Sigler},
  \bibinfo{person}{Mateusz Litwin}, \bibinfo{person}{Scott Gray},
  \bibinfo{person}{Benjamin Chess}, \bibinfo{person}{Jack Clark},
  \bibinfo{person}{Christopher Berner}, \bibinfo{person}{Sam McCandlish},
  \bibinfo{person}{Alec Radford}, \bibinfo{person}{Ilya Sutskever}, {and}
  \bibinfo{person}{Dario Amodei}.} \bibinfo{year}{2020}\natexlab{}.
\newblock \showarticletitle{Language Models are Few-Shot Learners}. In
  \bibinfo{booktitle}{\emph{Advances in Neural Information Processing
  Systems}}, \bibfield{editor}{\bibinfo{person}{H.~Larochelle},
  \bibinfo{person}{M.~Ranzato}, \bibinfo{person}{R.~Hadsell},
  \bibinfo{person}{M.~F. Balcan}, {and} \bibinfo{person}{H.~Lin}} (Eds.),
  Vol.~\bibinfo{volume}{33}. \bibinfo{publisher}{Curran Associates, Inc.},
  \bibinfo{pages}{1877--1901}.
\newblock
\urldef\tempurl%
\url{https://proceedings.neurips.cc/paper/2020/file/1457c0d6bfcb4967418bfb8ac142f64a-Paper.pdf}
\showURL{%
\tempurl}


\bibitem[\protect\citeauthoryear{Chakraborty, Ding, Allamanis, and
  Ray}{Chakraborty et~al\mbox{.}}{2020}]%
        {Chakraborty2020CODIT}
\bibfield{author}{\bibinfo{person}{Saikat Chakraborty},
  \bibinfo{person}{Yangruibo Ding}, \bibinfo{person}{Miltiadis Allamanis},
  {and} \bibinfo{person}{Baishakhi Ray}.} \bibinfo{year}{2020}\natexlab{}.
\newblock \showarticletitle{CODIT: Code Editing with Tree-Based Neural Models}.
\newblock \bibinfo{journal}{\emph{IEEE Transactions on Software Engineering}}
  (\bibinfo{year}{2020}), \bibinfo{pages}{1--1}.
\newblock
\showISSN{2326-3881}
\urldef\tempurl%
\url{https://doi.org/10.1109/tse.2020.3020502}
\showDOI{\tempurl}


\bibitem[\protect\citeauthoryear{Chen, Pei, and Furia}{Chen
  et~al\mbox{.}}{2017}]%
        {Chen2018JAID}
\bibfield{author}{\bibinfo{person}{Liushan Chen}, \bibinfo{person}{Yu Pei},
  {and} \bibinfo{person}{Carlo~A. Furia}.} \bibinfo{year}{2017}\natexlab{}.
\newblock \showarticletitle{Contract-based program repair without the
  contracts}. In \bibinfo{booktitle}{\emph{2017 32nd IEEE/ACM International
  Conference on Automated Software Engineering (ASE)}}.
  \bibinfo{pages}{637--647}.
\newblock
\urldef\tempurl%
\url{https://doi.org/10.1109/ASE.2017.8115674}
\showDOI{\tempurl}


\bibitem[\protect\citeauthoryear{Chen, Tworek, Jun, Yuan, de~Oliveira~Pinto,
  Kaplan, Edwards, Burda, Joseph, Brockman, Ray, Puri, Krueger, Petrov, Khlaaf,
  Sastry, Mishkin, Chan, Gray, Ryder, Pavlov, Power, Kaiser, Bavarian, Winter,
  Tillet, Such, Cummings, Plappert, Chantzis, Barnes, Herbert-Voss, Guss,
  Nichol, Paino, Tezak, Tang, Babuschkin, Balaji, Jain, Saunders, Hesse, Carr,
  Leike, Achiam, Misra, Morikawa, Radford, Knight, Brundage, Murati, Mayer,
  Welinder, McGrew, Amodei, McCandlish, Sutskever, and Zaremba}{Chen
  et~al\mbox{.}}{2021}]%
        {chen2021codex}
\bibfield{author}{\bibinfo{person}{Mark Chen}, \bibinfo{person}{Jerry Tworek},
  \bibinfo{person}{Heewoo Jun}, \bibinfo{person}{Qiming Yuan},
  \bibinfo{person}{Henrique~Ponde de Oliveira~Pinto}, \bibinfo{person}{Jared
  Kaplan}, \bibinfo{person}{Harri Edwards}, \bibinfo{person}{Yuri Burda},
  \bibinfo{person}{Nicholas Joseph}, \bibinfo{person}{Greg Brockman},
  \bibinfo{person}{Alex Ray}, \bibinfo{person}{Raul Puri},
  \bibinfo{person}{Gretchen Krueger}, \bibinfo{person}{Michael Petrov},
  \bibinfo{person}{Heidy Khlaaf}, \bibinfo{person}{Girish Sastry},
  \bibinfo{person}{Pamela Mishkin}, \bibinfo{person}{Brooke Chan},
  \bibinfo{person}{Scott Gray}, \bibinfo{person}{Nick Ryder},
  \bibinfo{person}{Mikhail Pavlov}, \bibinfo{person}{Alethea Power},
  \bibinfo{person}{Lukasz Kaiser}, \bibinfo{person}{Mohammad Bavarian},
  \bibinfo{person}{Clemens Winter}, \bibinfo{person}{Philippe Tillet},
  \bibinfo{person}{Felipe~Petroski Such}, \bibinfo{person}{Dave Cummings},
  \bibinfo{person}{Matthias Plappert}, \bibinfo{person}{Fotios Chantzis},
  \bibinfo{person}{Elizabeth Barnes}, \bibinfo{person}{Ariel Herbert-Voss},
  \bibinfo{person}{William~Hebgen Guss}, \bibinfo{person}{Alex Nichol},
  \bibinfo{person}{Alex Paino}, \bibinfo{person}{Nikolas Tezak},
  \bibinfo{person}{Jie Tang}, \bibinfo{person}{Igor Babuschkin},
  \bibinfo{person}{Suchir Balaji}, \bibinfo{person}{Shantanu Jain},
  \bibinfo{person}{William Saunders}, \bibinfo{person}{Christopher Hesse},
  \bibinfo{person}{Andrew~N. Carr}, \bibinfo{person}{Jan Leike},
  \bibinfo{person}{Josh Achiam}, \bibinfo{person}{Vedant Misra},
  \bibinfo{person}{Evan Morikawa}, \bibinfo{person}{Alec Radford},
  \bibinfo{person}{Matthew Knight}, \bibinfo{person}{Miles Brundage},
  \bibinfo{person}{Mira Murati}, \bibinfo{person}{Katie Mayer},
  \bibinfo{person}{Peter Welinder}, \bibinfo{person}{Bob McGrew},
  \bibinfo{person}{Dario Amodei}, \bibinfo{person}{Sam McCandlish},
  \bibinfo{person}{Ilya Sutskever}, {and} \bibinfo{person}{Wojciech Zaremba}.}
  \bibinfo{year}{2021}\natexlab{}.
\newblock \bibinfo{title}{Evaluating Large Language Models Trained on Code}.
\newblock
\newblock
\showeprint[arxiv]{2107.03374}~[cs.LG]


\bibitem[\protect\citeauthoryear{Chen, Kommrusch, Tufano, Pouchet, Poshyvanyk,
  and Martin}{Chen et~al\mbox{.}}{2019}]%
        {Chen2019SequenceRSL}
\bibfield{author}{\bibinfo{person}{Zimin Chen}, \bibinfo{person}{Steve
  Kommrusch}, \bibinfo{person}{Michele Tufano}, \bibinfo{person}{L. Pouchet},
  \bibinfo{person}{D. Poshyvanyk}, {and} \bibinfo{person}{Monperrus Martin}.}
  \bibinfo{year}{2019}\natexlab{}.
\newblock \showarticletitle{SequenceR: Sequence-to-Sequence Learning for
  End-to-End Program Repair}.
\newblock \bibinfo{journal}{\emph{ArXiv}}  \bibinfo{volume}{abs/1901.01808}
  (\bibinfo{year}{2019}).
\newblock


\bibitem[\protect\citeauthoryear{Cho, Merrienboer, \c{C}aglar G{\"u}l\c{c}ehre,
  Bahdanau, Bougares, Schwenk, and Bengio}{Cho et~al\mbox{.}}{2014}]%
        {Cho2014LearningPR}
\bibfield{author}{\bibinfo{person}{Kyunghyun Cho}, \bibinfo{person}{B.~V.
  Merrienboer}, \bibinfo{person}{\c{C}aglar G{\"u}l\c{c}ehre},
  \bibinfo{person}{Dzmitry Bahdanau}, \bibinfo{person}{Fethi Bougares},
  \bibinfo{person}{Holger Schwenk}, {and} \bibinfo{person}{Yoshua Bengio}.}
  \bibinfo{year}{2014}\natexlab{}.
\newblock \showarticletitle{Learning Phrase Representations using RNN
  Encoder-Decoder for Statistical Machine Translation}. In
  \bibinfo{booktitle}{\emph{EMNLP}}.
\newblock


\bibitem[\protect\citeauthoryear{Chung, \c{C}aglar G{\"u}l\c{c}ehre, Cho, and
  Bengio}{Chung et~al\mbox{.}}{2014}]%
        {Chung2014EmpiricalEO}
\bibfield{author}{\bibinfo{person}{J. Chung}, \bibinfo{person}{\c{C}aglar
  G{\"u}l\c{c}ehre}, \bibinfo{person}{Kyunghyun Cho}, {and}
  \bibinfo{person}{Yoshua Bengio}.} \bibinfo{year}{2014}\natexlab{}.
\newblock \showarticletitle{Empirical Evaluation of Gated Recurrent Neural
  Networks on Sequence Modeling}.
\newblock \bibinfo{journal}{\emph{ArXiv}}  \bibinfo{volume}{abs/1412.3555}
  (\bibinfo{year}{2014}).
\newblock


\bibitem[\protect\citeauthoryear{Cornu, Durieux, Seinturier, and
  Monperrus}{Cornu et~al\mbox{.}}{2015}]%
        {Cornu2015NPEFix}
\bibfield{author}{\bibinfo{person}{Benoit Cornu}, \bibinfo{person}{Thomas
  Durieux}, \bibinfo{person}{Lionel Seinturier}, {and} \bibinfo{person}{Martin
  Monperrus}.} \bibinfo{year}{2015}\natexlab{}.
\newblock \showarticletitle{NPEFix: Automatic Runtime Repair of Null Pointer
  Exceptions in Java}.
\newblock \bibinfo{journal}{\emph{CoRR}}  \bibinfo{volume}{abs/1512.07423}
  (\bibinfo{year}{2015}).
\newblock
\showeprint[arxiv]{1512.07423}
\urldef\tempurl%
\url{http://arxiv.org/abs/1512.07423}
\showURL{%
\tempurl}


\bibitem[\protect\citeauthoryear{DeMarco, Xuan, Le~Berre, and
  Monperrus}{DeMarco et~al\mbox{.}}{2014}]%
        {DeMarco2014ar}
\bibfield{author}{\bibinfo{person}{Favio DeMarco}, \bibinfo{person}{Jifeng
  Xuan}, \bibinfo{person}{Daniel Le~Berre}, {and} \bibinfo{person}{Martin
  Monperrus}.} \bibinfo{year}{2014}\natexlab{}.
\newblock \showarticletitle{Automatic Repair of Buggy If Conditions and Missing
  Preconditions with SMT}. In \bibinfo{booktitle}{\emph{Proceedings of the 6th
  International Workshop on Constraints in Software Testing, Verification, and
  Analysis}} (Hyderabad, India) \emph{(\bibinfo{series}{CSTVA 2014})}.
  \bibinfo{publisher}{Association for Computing Machinery},
  \bibinfo{address}{New York, NY, USA}, \bibinfo{pages}{30--39}.
\newblock
\showISBNx{9781450328470}
\urldef\tempurl%
\url{https://doi.org/10.1145/2593735.2593740}
\showDOI{\tempurl}


\bibitem[\protect\citeauthoryear{Durieux and Monperrus}{Durieux and
  Monperrus}{2016}]%
        {Durieux2016dynamoth}
\bibfield{author}{\bibinfo{person}{Thomas Durieux} {and}
  \bibinfo{person}{Martin Monperrus}.} \bibinfo{year}{2016}\natexlab{}.
\newblock \showarticletitle{DynaMoth: Dynamic Code Synthesis for Automatic
  Program Repair}. In \bibinfo{booktitle}{\emph{2016 IEEE/ACM 11th
  International Workshop in Automation of Software Test (AST)}}.
  \bibinfo{pages}{85--91}.
\newblock
\urldef\tempurl%
\url{https://doi.org/10.1109/AST.2016.021}
\showDOI{\tempurl}


\bibitem[\protect\citeauthoryear{Galenson, Reames, Bodik, Hartmann, and
  Sen}{Galenson et~al\mbox{.}}{2014}]%
        {Galenson2014CodeHint}
\bibfield{author}{\bibinfo{person}{Joel Galenson}, \bibinfo{person}{Philip
  Reames}, \bibinfo{person}{Rastislav Bodik}, \bibinfo{person}{Bj\"{o}rn
  Hartmann}, {and} \bibinfo{person}{Koushik Sen}.}
  \bibinfo{year}{2014}\natexlab{}.
\newblock \showarticletitle{CodeHint: Dynamic and Interactive Synthesis of Code
  Snippets}. In \bibinfo{booktitle}{\emph{Proceedings of the 36th International
  Conference on Software Engineering}} (Hyderabad, India)
  \emph{(\bibinfo{series}{ICSE 2014})}. \bibinfo{publisher}{Association for
  Computing Machinery}, \bibinfo{address}{New York, NY, USA},
  \bibinfo{pages}{653--663}.
\newblock
\showISBNx{9781450327565}
\urldef\tempurl%
\url{https://doi.org/10.1145/2568225.2568250}
\showDOI{\tempurl}


\bibitem[\protect\citeauthoryear{Gazzola, Micucci, and Mariani}{Gazzola
  et~al\mbox{.}}{2019}]%
        {Gazolla2019as}
\bibfield{author}{\bibinfo{person}{Luca Gazzola}, \bibinfo{person}{Daniela
  Micucci}, {and} \bibinfo{person}{Leonardo Mariani}.}
  \bibinfo{year}{2019}\natexlab{}.
\newblock \showarticletitle{Automatic Software Repair: A Survey}.
\newblock \bibinfo{journal}{\emph{IEEE Transactions on Software Engineering}}
  \bibinfo{volume}{45}, \bibinfo{number}{1} (\bibinfo{year}{2019}),
  \bibinfo{pages}{34--67}.
\newblock
\urldef\tempurl%
\url{https://doi.org/10.1109/TSE.2017.2755013}
\showDOI{\tempurl}


\bibitem[\protect\citeauthoryear{Ghanbari, Benton, and Zhang}{Ghanbari
  et~al\mbox{.}}{2019}]%
        {Ghanbari2019PraPR}
\bibfield{author}{\bibinfo{person}{Ali Ghanbari}, \bibinfo{person}{Samuel
  Benton}, {and} \bibinfo{person}{Lingming Zhang}.}
  \bibinfo{year}{2019}\natexlab{}.
\newblock \showarticletitle{Practical Program Repair via Bytecode Mutation}. In
  \bibinfo{booktitle}{\emph{Proceedings of the 28th ACM SIGSOFT International
  Symposium on Software Testing and Analysis}} (Beijing, China)
  \emph{(\bibinfo{series}{ISSTA 2019})}. \bibinfo{publisher}{Association for
  Computing Machinery}, \bibinfo{address}{New York, NY, USA},
  \bibinfo{pages}{19--30}.
\newblock
\showISBNx{9781450362245}
\urldef\tempurl%
\url{https://doi.org/10.1145/3293882.3330559}
\showDOI{\tempurl}


\bibitem[\protect\citeauthoryear{Hata, Shihab, and Neubig}{Hata
  et~al\mbox{.}}{2018a}]%
        {Hata2018LearningTG}
\bibfield{author}{\bibinfo{person}{Hideaki Hata}, \bibinfo{person}{Emad
  Shihab}, {and} \bibinfo{person}{Graham Neubig}.}
  \bibinfo{year}{2018}\natexlab{a}.
\newblock \showarticletitle{Learning to Generate Corrective Patches using
  Neural Machine Translation}.
\newblock \bibinfo{journal}{\emph{ArXiv}}  \bibinfo{volume}{abs/1812.07170}
  (\bibinfo{year}{2018}).
\newblock


\bibitem[\protect\citeauthoryear{Hata, Shihab, and Neubig}{Hata
  et~al\mbox{.}}{2018b}]%
        {Hata2018Ratchet}
\bibfield{author}{\bibinfo{person}{Hideaki Hata}, \bibinfo{person}{Emad
  Shihab}, {and} \bibinfo{person}{Graham Neubig}.}
  \bibinfo{year}{2018}\natexlab{b}.
\newblock \showarticletitle{Learning to Generate Corrective Patches using
  Neural Machine Translation}.
\newblock \bibinfo{journal}{\emph{CoRR}}  \bibinfo{volume}{abs/1812.07170}
  (\bibinfo{year}{2018}).
\newblock
\showeprint[arxiv]{1812.07170}
\urldef\tempurl%
\url{http://arxiv.org/abs/1812.07170}
\showURL{%
\tempurl}


\bibitem[\protect\citeauthoryear{Hindle, Barr, Su, Gabel, and Devanbu}{Hindle
  et~al\mbox{.}}{2012}]%
        {Hindle2012nat}
\bibfield{author}{\bibinfo{person}{Abram Hindle}, \bibinfo{person}{Earl~T.
  Barr}, \bibinfo{person}{Zhendong Su}, \bibinfo{person}{Mark Gabel}, {and}
  \bibinfo{person}{Premkumar Devanbu}.} \bibinfo{year}{2012}\natexlab{}.
\newblock \showarticletitle{On the Naturalness of Software}. In
  \bibinfo{booktitle}{\emph{Proceedings of the 34th International Conference on
  Software Engineering}} (Zurich, Switzerland) \emph{(\bibinfo{series}{ICSE
  '12})}. \bibinfo{publisher}{IEEE Press}, \bibinfo{pages}{837--847}.
\newblock
\showISBNx{9781467310673}


\bibitem[\protect\citeauthoryear{Hua, Zhang, Wang, and Khurshid}{Hua
  et~al\mbox{.}}{2018}]%
        {Hua2018SketchFix}
\bibfield{author}{\bibinfo{person}{Jinru Hua}, \bibinfo{person}{Mengshi Zhang},
  \bibinfo{person}{Kaiyuan Wang}, {and} \bibinfo{person}{Sarfraz Khurshid}.}
  \bibinfo{year}{2018}\natexlab{}.
\newblock \showarticletitle{SketchFix: A Tool for Automated Program Repair
  Approach Using Lazy Candidate Generation}. In
  \bibinfo{booktitle}{\emph{Proceedings of the 2018 26th ACM Joint Meeting on
  European Software Engineering Conference and Symposium on the Foundations of
  Software Engineering}} (Lake Buena Vista, FL, USA)
  \emph{(\bibinfo{series}{ESEC/FSE 2018})}. \bibinfo{publisher}{Association for
  Computing Machinery}, \bibinfo{address}{New York, NY, USA},
  \bibinfo{pages}{888--891}.
\newblock
\showISBNx{9781450355735}
\urldef\tempurl%
\url{https://doi.org/10.1145/3236024.3264600}
\showDOI{\tempurl}


\bibitem[\protect\citeauthoryear{Jiang, Ren, Xiong, and Zhang}{Jiang
  et~al\mbox{.}}{2019}]%
        {Jiang2019GenPat}
\bibfield{author}{\bibinfo{person}{Jiajun Jiang}, \bibinfo{person}{Luyao Ren},
  \bibinfo{person}{Yingfei Xiong}, {and} \bibinfo{person}{Lingming Zhang}.}
  \bibinfo{year}{2019}\natexlab{}.
\newblock \showarticletitle{Inferring Program Transformations From Singular
  Examples via Big Code}. In \bibinfo{booktitle}{\emph{2019 34th IEEE/ACM
  International Conference on Automated Software Engineering (ASE)}}.
  \bibinfo{pages}{255--266}.
\newblock
\urldef\tempurl%
\url{https://doi.org/10.1109/ASE.2019.00033}
\showDOI{\tempurl}


\bibitem[\protect\citeauthoryear{Jiang, Xiong, Zhang, Gao, and Chen}{Jiang
  et~al\mbox{.}}{2018}]%
        {Jiang2018ShapingPR}
\bibfield{author}{\bibinfo{person}{J. Jiang}, \bibinfo{person}{Yingfei Xiong},
  \bibinfo{person}{H. Zhang}, \bibinfo{person}{Q. Gao}, {and}
  \bibinfo{person}{X. Chen}.} \bibinfo{year}{2018}\natexlab{}.
\newblock \showarticletitle{Shaping program repair space with existing patches
  and similar code}.
\newblock \bibinfo{journal}{\emph{Proceedings of the 27th ACM SIGSOFT
  International Symposium on Software Testing and Analysis}}
  (\bibinfo{year}{2018}).
\newblock


\bibitem[\protect\citeauthoryear{Jiang, Lutellier, and Tan}{Jiang
  et~al\mbox{.}}{2021}]%
        {Lutellier2021ca}
\bibfield{author}{\bibinfo{person}{Nan Jiang}, \bibinfo{person}{Thibaud
  Lutellier}, {and} \bibinfo{person}{Lin Tan}.}
  \bibinfo{year}{2021}\natexlab{}.
\newblock \showarticletitle{CURE: Code-Aware Neural Machine Translation for
  Automatic Program Repair}. In \bibinfo{booktitle}{\emph{2021 IEEE/ACM 43rd
  International Conference on Software Engineering (ICSE)}}.
  \bibinfo{pages}{1161--1173}.
\newblock
\urldef\tempurl%
\url{https://doi.org/10.1109/ICSE43902.2021.00107}
\showDOI{\tempurl}


\bibitem[\protect\citeauthoryear{Just, Jalali, and Ernst}{Just
  et~al\mbox{.}}{2014}]%
        {Rene2014dj}
\bibfield{author}{\bibinfo{person}{Ren\'{e} Just}, \bibinfo{person}{Darioush
  Jalali}, {and} \bibinfo{person}{Michael~D. Ernst}.}
  \bibinfo{year}{2014}\natexlab{}.
\newblock \showarticletitle{Defects4J: A Database of Existing Faults to Enable
  Controlled Testing Studies for Java Programs}. In
  \bibinfo{booktitle}{\emph{Proceedings of the 2014 International Symposium on
  Software Testing and Analysis}} (San Jose, CA, USA)
  \emph{(\bibinfo{series}{ISSTA 2014})}. \bibinfo{publisher}{Association for
  Computing Machinery}, \bibinfo{address}{New York, NY, USA},
  \bibinfo{pages}{437--440}.
\newblock
\showISBNx{9781450326452}
\urldef\tempurl%
\url{https://doi.org/10.1145/2610384.2628055}
\showDOI{\tempurl}


\bibitem[\protect\citeauthoryear{Karampatsis, Babii, Robbes, Sutton, and
  Janes}{Karampatsis et~al\mbox{.}}{2020}]%
        {Karampatsis2020bcbv}
\bibfield{author}{\bibinfo{person}{Rafael-Michael Karampatsis},
  \bibinfo{person}{Hlib Babii}, \bibinfo{person}{Romain Robbes},
  \bibinfo{person}{Charles Sutton}, {and} \bibinfo{person}{Andrea Janes}.}
  \bibinfo{year}{2020}\natexlab{}.
\newblock \showarticletitle{Big Code != Big Vocabulary: Open-Vocabulary Models
  for Source Code}. In \bibinfo{booktitle}{\emph{Proceedings of the ACM/IEEE
  42nd International Conference on Software Engineering}} (Seoul, South Korea)
  \emph{(\bibinfo{series}{ICSE '20})}. \bibinfo{publisher}{Association for
  Computing Machinery}, \bibinfo{address}{New York, NY, USA},
  \bibinfo{pages}{1073--1085}.
\newblock
\showISBNx{9781450371216}
\urldef\tempurl%
\url{https://doi.org/10.1145/3377811.3380342}
\showDOI{\tempurl}


\bibitem[\protect\citeauthoryear{Kim, Nam, Song, and Kim}{Kim
  et~al\mbox{.}}{2013}]%
        {Kim2013par}
\bibfield{author}{\bibinfo{person}{Dongsun Kim}, \bibinfo{person}{Jaechang
  Nam}, \bibinfo{person}{Jaewoo Song}, {and} \bibinfo{person}{Sunghun Kim}.}
  \bibinfo{year}{2013}\natexlab{}.
\newblock \showarticletitle{Automatic patch generation learned from
  human-written patches}. In \bibinfo{booktitle}{\emph{2013 35th International
  Conference on Software Engineering (ICSE)}}. \bibinfo{pages}{802--811}.
\newblock
\urldef\tempurl%
\url{https://doi.org/10.1109/ICSE.2013.6606626}
\showDOI{\tempurl}


\bibitem[\protect\citeauthoryear{Kim and Kim}{Kim and Kim}{2019}]%
        {Kim2019ConFix}
\bibfield{author}{\bibinfo{person}{Jindae Kim} {and} \bibinfo{person}{Sunghun
  Kim}.} \bibinfo{year}{2019}\natexlab{}.
\newblock \showarticletitle{Automatic patch generation with context-based
  change application}.
\newblock \bibinfo{journal}{\emph{Empirical Software Engineering}}
  \bibinfo{volume}{24} (\bibinfo{date}{12} \bibinfo{year}{2019}).
\newblock
\urldef\tempurl%
\url{https://doi.org/10.1007/s10664-019-09742-5}
\showDOI{\tempurl}


\bibitem[\protect\citeauthoryear{Kim, Zhao, Tian, and Chandra}{Kim
  et~al\mbox{.}}{2021}]%
        {Kim2021CodePF}
\bibfield{author}{\bibinfo{person}{Seohyun Kim}, \bibinfo{person}{Jinman Zhao},
  \bibinfo{person}{Yuchi Tian}, {and} \bibinfo{person}{Satish Chandra}.}
  \bibinfo{year}{2021}\natexlab{}.
\newblock \showarticletitle{Code Prediction by Feeding Trees to Transformers}.
  In \bibinfo{booktitle}{\emph{2021 IEEE/ACM 43rd International Conference on
  Software Engineering (ICSE)}}. \bibinfo{pages}{150--162}.
\newblock
\urldef\tempurl%
\url{https://doi.org/10.1109/ICSE43902.2021.00026}
\showDOI{\tempurl}


\bibitem[\protect\citeauthoryear{Koyuncu, Liu, Bissyand{\'e}, Kim, Klein,
  Martin, and Traon}{Koyuncu et~al\mbox{.}}{2020}]%
        {Koyuncu2020FixMinerMR}
\bibfield{author}{\bibinfo{person}{A. Koyuncu}, \bibinfo{person}{K. Liu},
  \bibinfo{person}{Tegawend{\'e}~F. Bissyand{\'e}}, \bibinfo{person}{D. Kim},
  \bibinfo{person}{J. Klein}, \bibinfo{person}{Monperrus Martin}, {and}
  \bibinfo{person}{Y.~Le Traon}.} \bibinfo{year}{2020}\natexlab{}.
\newblock \showarticletitle{FixMiner: Mining relevant fix patterns for
  automated program repair}.
\newblock \bibinfo{journal}{\emph{Empirical Software Engineering}}
  \bibinfo{volume}{25} (\bibinfo{year}{2020}), \bibinfo{pages}{1980--2024}.
\newblock


\bibitem[\protect\citeauthoryear{Koyuncu, Liu, Bissyand\'{e}, Kim, Monperrus,
  Klein, and Le~Traon}{Koyuncu et~al\mbox{.}}{2019}]%
        {Koyuncu2019IFixR}
\bibfield{author}{\bibinfo{person}{Anil Koyuncu}, \bibinfo{person}{Kui Liu},
  \bibinfo{person}{Tegawend\'{e}~F. Bissyand\'{e}}, \bibinfo{person}{Dongsun
  Kim}, \bibinfo{person}{Martin Monperrus}, \bibinfo{person}{Jacques Klein},
  {and} \bibinfo{person}{Yves Le~Traon}.} \bibinfo{year}{2019}\natexlab{}.
\newblock \showarticletitle{IFixR: Bug Report Driven Program Repair}. In
  \bibinfo{booktitle}{\emph{Proceedings of the 2019 27th ACM Joint Meeting on
  European Software Engineering Conference and Symposium on the Foundations of
  Software Engineering}} (Tallinn, Estonia) \emph{(\bibinfo{series}{ESEC/FSE
  2019})}. \bibinfo{publisher}{Association for Computing Machinery},
  \bibinfo{address}{New York, NY, USA}, \bibinfo{pages}{314--325}.
\newblock
\showISBNx{9781450355728}
\urldef\tempurl%
\url{https://doi.org/10.1145/3338906.3338935}
\showDOI{\tempurl}


\bibitem[\protect\citeauthoryear{Le, Chu, Lo, Le~Goues, and Visser}{Le
  et~al\mbox{.}}{2017}]%
        {Le2017S3}
\bibfield{author}{\bibinfo{person}{Xuan-Bach~D. Le}, \bibinfo{person}{Duc-Hiep
  Chu}, \bibinfo{person}{David Lo}, \bibinfo{person}{Claire Le~Goues}, {and}
  \bibinfo{person}{Willem Visser}.} \bibinfo{year}{2017}\natexlab{}.
\newblock \showarticletitle{S3: Syntax- and Semantic-Guided Repair Synthesis
  via Programming by Examples}. In \bibinfo{booktitle}{\emph{Proceedings of the
  2017 11th Joint Meeting on Foundations of Software Engineering}} (Paderborn,
  Germany) \emph{(\bibinfo{series}{ESEC/FSE 2017})}.
  \bibinfo{publisher}{Association for Computing Machinery},
  \bibinfo{address}{New York, NY, USA}, \bibinfo{pages}{593--604}.
\newblock
\showISBNx{9781450351058}
\urldef\tempurl%
\url{https://doi.org/10.1145/3106237.3106309}
\showDOI{\tempurl}


\bibitem[\protect\citeauthoryear{Le, Lo, and Goues}{Le et~al\mbox{.}}{2016}]%
        {Le2016HistoryDP}
\bibfield{author}{\bibinfo{person}{Xuan-Bach~D. Le}, \bibinfo{person}{D. Lo},
  {and} \bibinfo{person}{Claire~Le Goues}.} \bibinfo{year}{2016}\natexlab{}.
\newblock \showarticletitle{History Driven Program Repair}.
\newblock \bibinfo{journal}{\emph{2016 IEEE 23rd International Conference on
  Software Analysis, Evolution, and Reengineering (SANER)}}
  \bibinfo{volume}{1} (\bibinfo{year}{2016}), \bibinfo{pages}{213--224}.
\newblock


\bibitem[\protect\citeauthoryear{{Le Goues}, {Nguyen}, {Forrest}, and
  {Weimer}}{{Le Goues} et~al\mbox{.}}{2012}]%
        {Goues2012tse}
\bibfield{author}{\bibinfo{person}{C. {Le Goues}}, \bibinfo{person}{T.
  {Nguyen}}, \bibinfo{person}{S. {Forrest}}, {and} \bibinfo{person}{W.
  {Weimer}}.} \bibinfo{year}{2012}\natexlab{}.
\newblock \showarticletitle{GenProg: A Generic Method for Automatic Software
  Repair}.
\newblock \bibinfo{journal}{\emph{IEEE Transactions on Software Engineering}}
  \bibinfo{volume}{38}, \bibinfo{number}{1} (\bibinfo{year}{2012}),
  \bibinfo{pages}{54--72}.
\newblock
\urldef\tempurl%
\url{https://doi.org/10.1109/TSE.2011.104}
\showDOI{\tempurl}


\bibitem[\protect\citeauthoryear{Li, Wang, and Nguyen}{Li
  et~al\mbox{.}}{2020}]%
        {Li2020DLFixCB}
\bibfield{author}{\bibinfo{person}{Yi Li}, \bibinfo{person}{Shaohua Wang},
  {and} \bibinfo{person}{Tien~N. Nguyen}.} \bibinfo{year}{2020}\natexlab{}.
\newblock \showarticletitle{DLFix: Context-Based Code Transformation Learning
  for Automated Program Repair}. In \bibinfo{booktitle}{\emph{Proceedings of
  the ACM/IEEE 42nd International Conference on Software Engineering}} (Seoul,
  South Korea) \emph{(\bibinfo{series}{ICSE '20})}.
  \bibinfo{publisher}{Association for Computing Machinery},
  \bibinfo{address}{New York, NY, USA}, \bibinfo{pages}{602--614}.
\newblock
\showISBNx{9781450371216}
\urldef\tempurl%
\url{https://doi.org/10.1145/3377811.3380345}
\showDOI{\tempurl}


\bibitem[\protect\citeauthoryear{Liu, Koyuncu, Bissyandé, Kim, Klein, and
  Le~Traon}{Liu et~al\mbox{.}}{2019a}]%
        {Liu2019kPAR}
\bibfield{author}{\bibinfo{person}{Kui Liu}, \bibinfo{person}{Anil Koyuncu},
  \bibinfo{person}{Tegawendé~F. Bissyandé}, \bibinfo{person}{Dongsun Kim},
  \bibinfo{person}{Jacques Klein}, {and} \bibinfo{person}{Yves Le~Traon}.}
  \bibinfo{year}{2019}\natexlab{a}.
\newblock \showarticletitle{You Cannot Fix What You Cannot Find! An
  Investigation of Fault Localization Bias in Benchmarking Automated Program
  Repair Systems}. In \bibinfo{booktitle}{\emph{2019 12th IEEE Conference on
  Software Testing, Validation and Verification (ICST)}}.
  \bibinfo{pages}{102--113}.
\newblock
\urldef\tempurl%
\url{https://doi.org/10.1109/ICST.2019.00020}
\showDOI{\tempurl}


\bibitem[\protect\citeauthoryear{Liu, Koyuncu, Kim, and Bissyand{\'e}}{Liu
  et~al\mbox{.}}{2019b}]%
        {Liu2019AVATARFS}
\bibfield{author}{\bibinfo{person}{K. Liu}, \bibinfo{person}{A. Koyuncu},
  \bibinfo{person}{D. Kim}, {and} \bibinfo{person}{Tegawend{\'e}~F.
  Bissyand{\'e}}.} \bibinfo{year}{2019}\natexlab{b}.
\newblock \showarticletitle{AVATAR: Fixing Semantic Bugs with Fix Patterns of
  Static Analysis Violations}.
\newblock \bibinfo{journal}{\emph{2019 IEEE 26th International Conference on
  Software Analysis, Evolution and Reengineering (SANER)}}
  (\bibinfo{year}{2019}), \bibinfo{pages}{1--12}.
\newblock


\bibitem[\protect\citeauthoryear{Liu, Koyuncu, Kim, and Bissyand\'{e}}{Liu
  et~al\mbox{.}}{2019c}]%
        {Liu2019tbar}
\bibfield{author}{\bibinfo{person}{Kui Liu}, \bibinfo{person}{Anil Koyuncu},
  \bibinfo{person}{Dongsun Kim}, {and} \bibinfo{person}{Tegawend\'{e}~F.
  Bissyand\'{e}}.} \bibinfo{year}{2019}\natexlab{c}.
\newblock \showarticletitle{TBar: Revisiting Template-Based Automated Program
  Repair} \emph{(\bibinfo{series}{ISSTA 2019})}.
  \bibinfo{publisher}{Association for Computing Machinery},
  \bibinfo{address}{New York, NY, USA}, \bibinfo{pages}{31--42}.
\newblock
\showISBNx{9781450362245}
\urldef\tempurl%
\url{https://doi.org/10.1145/3293882.3330577}
\showDOI{\tempurl}


\bibitem[\protect\citeauthoryear{Liu, Koyuncu, Kim, Kim, and F.~Bissyandé}{Liu
  et~al\mbox{.}}{2018}]%
        {Liu2018LSRepair}
\bibfield{author}{\bibinfo{person}{Kui Liu}, \bibinfo{person}{Anil Koyuncu},
  \bibinfo{person}{Kisub Kim}, \bibinfo{person}{Dongsun Kim}, {and}
  \bibinfo{person}{Tegawendé F.~Bissyandé}.} \bibinfo{year}{2018}\natexlab{}.
\newblock \showarticletitle{LSRepair: Live Search of Fix Ingredients for
  Automated Program Repair}. In \bibinfo{booktitle}{\emph{2018 25th
  Asia-Pacific Software Engineering Conference (APSEC)}}.
  \bibinfo{pages}{658--662}.
\newblock
\urldef\tempurl%
\url{https://doi.org/10.1109/APSEC.2018.00085}
\showDOI{\tempurl}


\bibitem[\protect\citeauthoryear{Liu, Wang, Koyuncu, Kim, Bissyand\'{e}, Kim,
  Wu, Klein, Mao, and Traon}{Liu et~al\mbox{.}}{2020}]%
        {Liu2020ot}
\bibfield{author}{\bibinfo{person}{Kui Liu}, \bibinfo{person}{Shangwen Wang},
  \bibinfo{person}{Anil Koyuncu}, \bibinfo{person}{Kisub Kim},
  \bibinfo{person}{Tegawend\'{e}~F. Bissyand\'{e}}, \bibinfo{person}{Dongsun
  Kim}, \bibinfo{person}{Peng Wu}, \bibinfo{person}{Jacques Klein},
  \bibinfo{person}{Xiaoguang Mao}, {and} \bibinfo{person}{Yves~Le Traon}.}
  \bibinfo{year}{2020}\natexlab{}.
\newblock \showarticletitle{On the Efficiency of Test Suite Based Program
  Repair: A Systematic Assessment of 16 Automated Repair Systems for Java
  Programs}. In \bibinfo{booktitle}{\emph{Proceedings of the ACM/IEEE 42nd
  International Conference on Software Engineering}} (Seoul, South Korea)
  \emph{(\bibinfo{series}{ICSE '20})}. \bibinfo{publisher}{Association for
  Computing Machinery}, \bibinfo{address}{New York, NY, USA},
  \bibinfo{pages}{615--627}.
\newblock
\showISBNx{9781450371216}
\urldef\tempurl%
\url{https://doi.org/10.1145/3377811.3380338}
\showDOI{\tempurl}


\bibitem[\protect\citeauthoryear{Liu and Zhong}{Liu and Zhong}{2018}]%
        {Liu2018SOFix}
\bibfield{author}{\bibinfo{person}{Xuliang Liu} {and} \bibinfo{person}{Hao
  Zhong}.} \bibinfo{year}{2018}\natexlab{}.
\newblock \showarticletitle{Mining stackoverflow for program repair}. In
  \bibinfo{booktitle}{\emph{2018 IEEE 25th International Conference on Software
  Analysis, Evolution and Reengineering (SANER)}}. \bibinfo{pages}{118--129}.
\newblock
\urldef\tempurl%
\url{https://doi.org/10.1109/SANER.2018.8330202}
\showDOI{\tempurl}


\bibitem[\protect\citeauthoryear{Lutellier, Pham, Pang, Li, Wei, and
  Tan}{Lutellier et~al\mbox{.}}{2020}]%
        {Lutellier2020cc}
\bibfield{author}{\bibinfo{person}{Thibaud Lutellier},
  \bibinfo{person}{Hung~Viet Pham}, \bibinfo{person}{Lawrence Pang},
  \bibinfo{person}{Yitong Li}, \bibinfo{person}{Moshi Wei}, {and}
  \bibinfo{person}{Lin Tan}.} \bibinfo{year}{2020}\natexlab{}.
\newblock \showarticletitle{CoCoNuT: Combining Context-Aware Neural Translation
  Models Using Ensemble for Program Repair}. In
  \bibinfo{booktitle}{\emph{Proceedings of the 29th ACM SIGSOFT International
  Symposium on Software Testing and Analysis}} (Virtual Event, USA)
  \emph{(\bibinfo{series}{ISSTA 2020})}. \bibinfo{publisher}{Association for
  Computing Machinery}, \bibinfo{address}{New York, NY, USA},
  \bibinfo{pages}{101--114}.
\newblock
\showISBNx{9781450380089}
\urldef\tempurl%
\url{https://doi.org/10.1145/3395363.3397369}
\showDOI{\tempurl}


\bibitem[\protect\citeauthoryear{Martinez and Monperrus}{Martinez and
  Monperrus}{2016}]%
        {Martinez2016ASTOR}
\bibfield{author}{\bibinfo{person}{Matias Martinez} {and}
  \bibinfo{person}{Martin Monperrus}.} \bibinfo{year}{2016}\natexlab{}.
\newblock \showarticletitle{ASTOR: A Program Repair Library for Java (Demo)}.
  In \bibinfo{booktitle}{\emph{Proceedings of the 25th International Symposium
  on Software Testing and Analysis}} (Saarbr\"{u}cken, Germany)
  \emph{(\bibinfo{series}{ISSTA 2016})}. \bibinfo{publisher}{Association for
  Computing Machinery}, \bibinfo{address}{New York, NY, USA},
  \bibinfo{pages}{441--444}.
\newblock
\showISBNx{9781450343909}
\urldef\tempurl%
\url{https://doi.org/10.1145/2931037.2948705}
\showDOI{\tempurl}


\bibitem[\protect\citeauthoryear{Martinez and Monperrus}{Martinez and
  Monperrus}{2018}]%
        {Martinez2018Cardumen}
\bibfield{author}{\bibinfo{person}{Matias Martinez} {and}
  \bibinfo{person}{Martin Monperrus}.} \bibinfo{year}{2018}\natexlab{}.
\newblock \bibinfo{booktitle}{\emph{Ultra-Large Repair Search Space with
  Automatically Mined Templates: The Cardumen Mode of Astor: 10th International
  Symposium, SSBSE 2018, Montpellier, France, September 8-9, 2018,
  Proceedings}}.
\newblock \bibinfo{pages}{65--86}.
\newblock
\showISBNx{978-3-319-99240-2}
\urldef\tempurl%
\url{https://doi.org/10.1007/978-3-319-99241-9_3}
\showDOI{\tempurl}


\bibitem[\protect\citeauthoryear{Mashhadi and Hemmati}{Mashhadi and
  Hemmati}{2021}]%
        {Mashhadi2021BERTAPR}
\bibfield{author}{\bibinfo{person}{E. Mashhadi} {and} \bibinfo{person}{H.
  Hemmati}.} \bibinfo{year}{2021}\natexlab{}.
\newblock \showarticletitle{Applying CodeBERT for Automated Program Repair of
  Java Simple Bugs}. In \bibinfo{booktitle}{\emph{2021 2021 IEEE/ACM 18th
  International Conference on Mining Software Repositories (MSR) (MSR)}}.
  \bibinfo{publisher}{IEEE Computer Society}, \bibinfo{address}{Los Alamitos,
  CA, USA}, \bibinfo{pages}{505--509}.
\newblock
\urldef\tempurl%
\url{https://doi.org/10.1109/MSR52588.2021.00063}
\showDOI{\tempurl}


\bibitem[\protect\citeauthoryear{Monperrus}{Monperrus}{2020}]%
        {monperrus2020livingreview}
\bibfield{author}{\bibinfo{person}{Martin Monperrus}.}
  \bibinfo{year}{2020}\natexlab{}.
\newblock \bibinfo{title}{{The Living Review on Automated Program Repair}}.
  (\bibinfo{date}{Dec.} \bibinfo{year}{2020}).
\newblock
\urldef\tempurl%
\url{https://hal.archives-ouvertes.fr/hal-01956501}
\showURL{%
\tempurl}
\newblock
\shownote{working paper or preprint.}


\bibitem[\protect\citeauthoryear{Radford, Narasimhan, Salimans, and
  Sutskever}{Radford et~al\mbox{.}}{2018}]%
        {radford2018GPT1}
\bibfield{author}{\bibinfo{person}{Alec Radford}, \bibinfo{person}{Karthik
  Narasimhan}, \bibinfo{person}{Tim Salimans}, {and} \bibinfo{person}{Ilya
  Sutskever}.} \bibinfo{year}{2018}\natexlab{}.
\newblock \showarticletitle{Improving language understanding by generative
  pre-training}.
\newblock  (\bibinfo{year}{2018}).
\newblock


\bibitem[\protect\citeauthoryear{Ray, Hellendoorn, Godhane, Tu, Bacchelli, and
  Devanbu}{Ray et~al\mbox{.}}{2016}]%
        {Ray2016buggynat}
\bibfield{author}{\bibinfo{person}{Baishakhi Ray}, \bibinfo{person}{Vincent
  Hellendoorn}, \bibinfo{person}{Saheel Godhane}, \bibinfo{person}{Zhaopeng
  Tu}, \bibinfo{person}{Alberto Bacchelli}, {and} \bibinfo{person}{Premkumar
  Devanbu}.} \bibinfo{year}{2016}\natexlab{}.
\newblock \showarticletitle{On the "Naturalness" of Buggy Code}. In
  \bibinfo{booktitle}{\emph{Proceedings of the 38th International Conference on
  Software Engineering}} (Austin, Texas) \emph{(\bibinfo{series}{ICSE '16})}.
  \bibinfo{publisher}{Association for Computing Machinery},
  \bibinfo{address}{New York, NY, USA}, \bibinfo{pages}{428--439}.
\newblock
\showISBNx{9781450339001}
\urldef\tempurl%
\url{https://doi.org/10.1145/2884781.2884848}
\showDOI{\tempurl}


\bibitem[\protect\citeauthoryear{Saha, Lyu, Yoshida, and Prasad}{Saha
  et~al\mbox{.}}{2017}]%
        {Saha2017Elixir}
\bibfield{author}{\bibinfo{person}{Ripon~K. Saha}, \bibinfo{person}{Yingjun
  Lyu}, \bibinfo{person}{Hiroaki Yoshida}, {and} \bibinfo{person}{Mukul~R.
  Prasad}.} \bibinfo{year}{2017}\natexlab{}.
\newblock \showarticletitle{Elixir: Effective object-oriented program repair}.
  In \bibinfo{booktitle}{\emph{2017 32nd IEEE/ACM International Conference on
  Automated Software Engineering (ASE)}}. \bibinfo{pages}{648--659}.
\newblock
\urldef\tempurl%
\url{https://doi.org/10.1109/ASE.2017.8115675}
\showDOI{\tempurl}


\bibitem[\protect\citeauthoryear{Saha, Saha, and Prasad}{Saha
  et~al\mbox{.}}{2019}]%
        {Saha2019HarnessingEF}
\bibfield{author}{\bibinfo{person}{Seemanta Saha}, \bibinfo{person}{R. Saha},
  {and} \bibinfo{person}{M. Prasad}.} \bibinfo{year}{2019}\natexlab{}.
\newblock \showarticletitle{Harnessing Evolution for Multi-Hunk Program
  Repair}.
\newblock \bibinfo{journal}{\emph{2019 IEEE/ACM 41st International Conference
  on Software Engineering (ICSE)}} (\bibinfo{year}{2019}),
  \bibinfo{pages}{13--24}.
\newblock


\bibitem[\protect\citeauthoryear{Sobreira, Durieux, Madeiral, Monperrus, and
  Maia}{Sobreira et~al\mbox{.}}{2018}]%
        {Sobreira2018dissection}
\bibfield{author}{\bibinfo{person}{Victor Sobreira}, \bibinfo{person}{Thomas
  Durieux}, \bibinfo{person}{Fernanda Madeiral}, \bibinfo{person}{Martin
  Monperrus}, {and} \bibinfo{person}{Marcelo~A. Maia}.}
  \bibinfo{year}{2018}\natexlab{}.
\newblock \showarticletitle{{Dissection of a Bug Dataset: Anatomy of 395
  Patches from Defects4J}}. In \bibinfo{booktitle}{\emph{Proceedings of
  SANER}}.
\newblock


\bibitem[\protect\citeauthoryear{Sutskever, Vinyals, and Le}{Sutskever
  et~al\mbox{.}}{2014}]%
        {Sutskever2014SequenceTS}
\bibfield{author}{\bibinfo{person}{Ilya Sutskever}, \bibinfo{person}{Oriol
  Vinyals}, {and} \bibinfo{person}{Quoc~V. Le}.}
  \bibinfo{year}{2014}\natexlab{}.
\newblock \showarticletitle{Sequence to Sequence Learning with Neural
  Networks}. In \bibinfo{booktitle}{\emph{NIPS}}.
\newblock


\bibitem[\protect\citeauthoryear{Wang, Meng, Wang, Liu, and Hao}{Wang
  et~al\mbox{.}}{2019}]%
        {Wang2019LoopFix}
\bibfield{author}{\bibinfo{person}{Weichao Wang}, \bibinfo{person}{Zhaopeng
  Meng}, \bibinfo{person}{Zan Wang}, \bibinfo{person}{Shuang Liu}, {and}
  \bibinfo{person}{Jianye Hao}.} \bibinfo{year}{2019}\natexlab{}.
\newblock \showarticletitle{LoopFix: an approach to automatic repair of buggy
  loops}.
\newblock \bibinfo{journal}{\emph{Journal of Systems and Software}}
  \bibinfo{volume}{156} (\bibinfo{year}{2019}), \bibinfo{pages}{100--112}.
\newblock
\showISSN{0164-1212}
\urldef\tempurl%
\url{https://doi.org/10.1016/j.jss.2019.06.076}
\showDOI{\tempurl}


\bibitem[\protect\citeauthoryear{Wen, Chen, Wu, Hao, and Cheung}{Wen
  et~al\mbox{.}}{2018}]%
        {Wen2018CapGen}
\bibfield{author}{\bibinfo{person}{Ming Wen}, \bibinfo{person}{Junjie Chen},
  \bibinfo{person}{Rongxin Wu}, \bibinfo{person}{Dan Hao}, {and}
  \bibinfo{person}{Shing-Chi Cheung}.} \bibinfo{year}{2018}\natexlab{}.
\newblock \showarticletitle{Context-Aware Patch Generation for Better Automated
  Program Repair}. In \bibinfo{booktitle}{\emph{Proceedings of the 40th
  International Conference on Software Engineering}} (Gothenburg, Sweden)
  \emph{(\bibinfo{series}{ICSE '18})}. \bibinfo{publisher}{Association for
  Computing Machinery}, \bibinfo{address}{New York, NY, USA},
  \bibinfo{pages}{1--11}.
\newblock
\showISBNx{9781450356381}
\urldef\tempurl%
\url{https://doi.org/10.1145/3180155.3180233}
\showDOI{\tempurl}


\bibitem[\protect\citeauthoryear{White, Tufano, Martínez, Monperrus, and
  Poshyvanyk}{White et~al\mbox{.}}{2019}]%
        {White2019DeepRepair}
\bibfield{author}{\bibinfo{person}{Martin White}, \bibinfo{person}{Michele
  Tufano}, \bibinfo{person}{Matías Martínez}, \bibinfo{person}{Martin
  Monperrus}, {and} \bibinfo{person}{Denys Poshyvanyk}.}
  \bibinfo{year}{2019}\natexlab{}.
\newblock \showarticletitle{Sorting and Transforming Program Repair Ingredients
  via Deep Learning Code Similarities}. In \bibinfo{booktitle}{\emph{2019 IEEE
  26th International Conference on Software Analysis, Evolution and
  Reengineering (SANER)}}. \bibinfo{pages}{479--490}.
\newblock
\urldef\tempurl%
\url{https://doi.org/10.1109/SANER.2019.8668043}
\showDOI{\tempurl}


\bibitem[\protect\citeauthoryear{Wong, Santiesteban, K\"{a}stner, and
  Le~Goues}{Wong et~al\mbox{.}}{2021}]%
        {Wong2021VarFix}
\bibfield{author}{\bibinfo{person}{Chu-Pan Wong}, \bibinfo{person}{Priscila
  Santiesteban}, \bibinfo{person}{Christian K\"{a}stner}, {and}
  \bibinfo{person}{Claire Le~Goues}.} \bibinfo{year}{2021}\natexlab{}.
\newblock \bibinfo{booktitle}{\emph{VarFix: Balancing Edit Expressiveness and
  Search Effectiveness in Automated Program Repair}}.
\newblock \bibinfo{publisher}{Association for Computing Machinery},
  \bibinfo{address}{New York, NY, USA}, \bibinfo{pages}{354--366}.
\newblock
\showISBNx{9781450385626}
\urldef\tempurl%
\url{https://doi.org/10.1145/3468264.3468600}
\showURL{%
\tempurl}


\bibitem[\protect\citeauthoryear{Xin and Reiss}{Xin and Reiss}{2017}]%
        {Xin2017ssFix}
\bibfield{author}{\bibinfo{person}{Qi Xin} {and} \bibinfo{person}{Steven~P.
  Reiss}.} \bibinfo{year}{2017}\natexlab{}.
\newblock \showarticletitle{Leveraging syntax-related code for automated
  program repair}. In \bibinfo{booktitle}{\emph{2017 32nd IEEE/ACM
  International Conference on Automated Software Engineering (ASE)}}.
  \bibinfo{pages}{660--670}.
\newblock
\urldef\tempurl%
\url{https://doi.org/10.1109/ASE.2017.8115676}
\showDOI{\tempurl}


\bibitem[\protect\citeauthoryear{Xiong, Wang, Yan, Zhang, Han, Huang, and
  Zhang}{Xiong et~al\mbox{.}}{2017}]%
        {Xiong2017acs}
\bibfield{author}{\bibinfo{person}{Yingfei Xiong}, \bibinfo{person}{Jie Wang},
  \bibinfo{person}{Runfa Yan}, \bibinfo{person}{Jiachen Zhang},
  \bibinfo{person}{Shi Han}, \bibinfo{person}{Gang Huang}, {and}
  \bibinfo{person}{Lu Zhang}.} \bibinfo{year}{2017}\natexlab{}.
\newblock \showarticletitle{Precise Condition Synthesis for Program Repair}. In
  \bibinfo{booktitle}{\emph{Proceedings of the 39th International Conference on
  Software Engineering}} (Buenos Aires, Argentina) \emph{(\bibinfo{series}{ICSE
  '17})}. \bibinfo{publisher}{IEEE Press}, \bibinfo{pages}{416--426}.
\newblock
\showISBNx{9781538638682}
\urldef\tempurl%
\url{https://doi.org/10.1109/ICSE.2017.45}
\showDOI{\tempurl}


\bibitem[\protect\citeauthoryear{Xu, Chen, Pei, Zhang, Pan, and Furia}{Xu
  et~al\mbox{.}}{5555}]%
        {Xu2020Restore}
\bibfield{author}{\bibinfo{person}{T. Xu}, \bibinfo{person}{L. Chen},
  \bibinfo{person}{Y. Pei}, \bibinfo{person}{T. Zhang}, \bibinfo{person}{M.
  Pan}, {and} \bibinfo{person}{C. Furia}.} \bibinfo{year}{5555}\natexlab{}.
\newblock \showarticletitle{Restore: Retrospective Fault Localization Enhancing
  Automated Program Repair}.
\newblock \bibinfo{journal}{\emph{IEEE Transactions on Software Engineering}}
  \bibinfo{number}{01} (\bibinfo{date}{apr} \bibinfo{year}{5555}),
  \bibinfo{pages}{1--1}.
\newblock
\showISSN{1939-3520}
\urldef\tempurl%
\url{https://doi.org/10.1109/TSE.2020.2987862}
\showDOI{\tempurl}


\bibitem[\protect\citeauthoryear{Xu, Sui, Yan, and Xue}{Xu
  et~al\mbox{.}}{2019}]%
        {Xu2019VFix}
\bibfield{author}{\bibinfo{person}{Xuezheng Xu}, \bibinfo{person}{Yulei Sui},
  \bibinfo{person}{Hua Yan}, {and} \bibinfo{person}{Jingling Xue}.}
  \bibinfo{year}{2019}\natexlab{}.
\newblock \showarticletitle{VFix: Value-Flow-Guided Precise Program Repair for
  Null Pointer Dereferences}. In \bibinfo{booktitle}{\emph{Proceedings of the
  41st International Conference on Software Engineering}} (Montreal, Quebec,
  Canada) \emph{(\bibinfo{series}{ICSE '19})}. \bibinfo{publisher}{IEEE Press},
  \bibinfo{pages}{512--523}.
\newblock
\urldef\tempurl%
\url{https://doi.org/10.1109/ICSE.2019.00063}
\showDOI{\tempurl}


\bibitem[\protect\citeauthoryear{Yuan and Banzhaf}{Yuan and Banzhaf}{2020}]%
        {Yuan2020ARJA}
\bibfield{author}{\bibinfo{person}{Yuan Yuan} {and} \bibinfo{person}{Wolfgang
  Banzhaf}.} \bibinfo{year}{2020}\natexlab{}.
\newblock \showarticletitle{Toward Better Evolutionary Program Repair: An
  Integrated Approach}.
\newblock  \bibinfo{volume}{29}, \bibinfo{number}{1}, Article
  \bibinfo{articleno}{5} (\bibinfo{date}{Jan.} \bibinfo{year}{2020}),
  \bibinfo{numpages}{53}~pages.
\newblock
\showISSN{1049-331X}
\urldef\tempurl%
\url{https://doi.org/10.1145/3360004}
\showDOI{\tempurl}


\bibitem[\protect\citeauthoryear{Zeller}{Zeller}{2002}]%
        {Zeller2002ic}
\bibfield{author}{\bibinfo{person}{Andreas Zeller}.}
  \bibinfo{year}{2002}\natexlab{}.
\newblock \showarticletitle{Isolating Cause-Effect Chains from Computer
  Programs}. In \bibinfo{booktitle}{\emph{Proceedings of the 10th ACM SIGSOFT
  Symposium on Foundations of Software Engineering}} (Charleston, South
  Carolina, USA) \emph{(\bibinfo{series}{SIGSOFT '02/FSE-10})}.
  \bibinfo{publisher}{Association for Computing Machinery},
  \bibinfo{address}{New York, NY, USA}, \bibinfo{pages}{1--10}.
\newblock
\showISBNx{1581135149}
\urldef\tempurl%
\url{https://doi.org/10.1145/587051.587053}
\showDOI{\tempurl}


\bibitem[\protect\citeauthoryear{Zhu, Sun, Xiao, Zhang, Yuan, Xiong, and
  Zhang}{Zhu et~al\mbox{.}}{2021}]%
        {Zhu2021Recoder}
\bibfield{author}{\bibinfo{person}{Qihao Zhu}, \bibinfo{person}{Zeyu Sun},
  \bibinfo{person}{Yuan-an Xiao}, \bibinfo{person}{Wenjie Zhang},
  \bibinfo{person}{Kang Yuan}, \bibinfo{person}{Yingfei Xiong}, {and}
  \bibinfo{person}{Lu Zhang}.} \bibinfo{year}{2021}\natexlab{}.
\newblock \bibinfo{booktitle}{\emph{A Syntax-Guided Edit Decoder for Neural
  Program Repair}}.
\newblock \bibinfo{publisher}{Association for Computing Machinery},
  \bibinfo{address}{New York, NY, USA}, \bibinfo{pages}{341--353}.
\newblock
\showISBNx{9781450385626}
\urldef\tempurl%
\url{https://doi.org/10.1145/3468264.3468544}
\showURL{%
\tempurl}


\end{thebibliography}

\appendix

\end{document}
\endinput
%%
%% End of file `sample-sigconf.tex'.

\title{Appendix of \name: Predicate Synthesis with Language Models}

\maketitle

\section{Source for Existing Work Results}

We present the source for our existing work results in Table~\ref{tab:result_source}, as well as publications for which we could not find bug-level results.

\begin{table*}[ht]
    \centering
    \caption{Result sources for \dfj bugs that each tool fixes. \label{tab:result_source}}
    \scalebox{1.0}{
    \begin{tabular}{p{0.05\textwidth}p{0.1\textwidth}p{0.05\textwidth}p{0.17\textwidth}p{0.53\textwidth}}
    \toprule
    No. & Tool Name & Pub. & Source of Results & Notes \\\midrule
    1 & $TBar_p$ & \cite{Liu2019tbar} & \href{https://github.com/TruX-DTF/TBar/tree/master/Results/PerfectFL/TBar/FixedBugs}{Github link} & Bugs without \texttt{\_P} suffix are regarded correct.\\
    2 & TBar-10k & \cite{Liu2020ot} & \href{https://github.com/TruX-DTF/APR-Efficiency/tree/master/Patches/PFL/TBar}{Github link} & Bugs with \texttt{\_C} suffix are regarded correct. \\
    3 & CURE & \cite{Lutellier2021ca} & \href{https://github.com/lin-tan/CURE/tree/a804ceee2d10028075aad851f34c9270cdd9a0ab/Bugs_CURE_fixed/Defects4J}{Github link} & We could not find which bugs were plausibly fixed by CURE. \\
    4 & HDRepair & \cite{Le2016HistoryDP} & \href{https://github.com/xuanbachle/bugfixes/blob/7ba0da0c554fd0e09a03e5c3bfa566160ec8b4dd/fixed.txt}{Github link} & - \\
    5 & S3 & \cite{Le2017S3} & \href{https://github.com/anonymousfserepair2017/FSE2017-S3-SyntaxSemanticRepairData}{Github link} & The repository linked does not clearly mark which bugs are fixed. \\
    6 & Elixir & \cite{Saha2017Elixir} & - & We could not find individual results within the paper; there is no public repository of the results we could find. \\
    7 & SequenceR & \cite{Chen2019SequenceRSL} & \href{https://github.com/KTH/chai/blob/c5c07e18a2f40728714d08a18bb5d69d18d1f746/results/Defects4J_patches/PassComments.txt}{Github link} & - \\
    8 & Hercules & \cite{Saha2019HerculesPerBugArxiv} & Table 4, Table 5 of paper & - \\
    9 & SimFix & \cite{Jiang2018ShapingPR} & \href{https://github.com/TruX-DTF/APR-Efficiency/tree/6bc7e2e0a8c5a096e96f8336d62daa6900b6474a/Patches/PFL/SimFix}{Github link} & Bugs with \texttt{\_C} suffix are regarded correct. \\
    10 & FixMiner & \cite{Koyuncu2020FixMinerMR} & \href{https://github.com/TruX-DTF/APR-Efficiency/tree/6bc7e2e0a8c5a096e96f8336d62daa6900b6474a/Patches/PFL/FixMiner}{Github link} & Bugs with \texttt{\_C} suffix are regarded correct. \\
    11 & DLFix & \cite{Li2020DLFixCB} & \href{https://github.com/ICSE-2019-AUTOFIX/ICSE-2019-AUTOFIX/tree/3d741fb03420ad20dea157a943df06d487723e99/results}{Github link} & We use the perfect FL assumption, so bugs marked with (Buggy position known) are treated as fixed. \\
    12 & ACS & \cite{Xiong2017acs} & \href{https://github.com/TruX-DTF/APR-Efficiency/tree/6bc7e2e0a8c5a096e96f8336d62daa6900b6474a/Patches/PFL/ACS}{Github link} & Bugs with \texttt{\_C} suffix are regarded correct. \\
    13 & CapGen & \cite{Wen2018CapGen} & Table 3 of paper & - \\
    14 & PAR & \cite{Kim2013par} & \href{https://sites.google.com/site/autofixhkust/home/patch-comparison}{Google sites link} & The results are reported by the original project's bug ID; we translate those to Defects4J bug ID. \\
    15 & SOFix & \cite{Liu2018SOFix} & Table VI of paper & - \\
    16 & PraPR & \cite{Ghanbari2019PraPR} & Table 5 of paper & Bugs colored dark gray on the All Mutators column are regarded fixed. \\
    17 & JAID & \cite{Chen2018JAID} & Table II of paper & Bugs in which the Fixes-Rank column does not have a hyphen are correctly fixed. \\
    18 & SketchFix & \cite{Hua2018SketchFix} & Figure 4 of paper & Bugs with colored rows are correctly fixed. \\
    19 & LSRepair & \cite{Liu2018LSRepair} & Table II of paper & - \\
    20 & ssFix & \cite{Xin2017ssFix} & \href{https://github.com/qixin5/ssFix/tree/2f6e0394c275866bf9437025f38e31683324cf11/expt0/patch/ssFix}{Github link} & Bugs noted as either Valid or Correct are regarded as correctly fixed. \\
    21 & DeepRepair & \cite{White2019DeepRepair} & \href{https://github.com/SpoonLabs/astor-experiments/tree/master/DeepRepair}{Github link} & The publication does not contain individual bug repair results; the link we provide does not make it clear which bugs were correctly fixed.\\
    22 & NPEfix & \cite{Cornu2015NPEFix} & \href{https://hal.archives-ouvertes.fr/hal-01419861/document}{Paper}, Table II & The results are reported by the original project's bug ID; we translate those to Defects4J bug ID. \\
    23 & Ratchet & \cite{Hata2018Ratchet} & Table 12 of paper & Closure-46 is marked as a complete (correct) fix. \\
    24 & Codit & \cite{Chakraborty2020CODIT} & Table 10 of paper & Rows marked with green are correct. \\
    25 & AVATAR & \cite{Liu2019AVATARFS} & \href{https://github.com/TruX-DTF/APR-Efficiency/tree/6bc7e2e0a8c5a096e96f8336d62daa6900b6474a/Patches/PFL/AVATAR}{Github link} & Bugs with \texttt{\_C} suffix are regarded correct. \\
    26 & CoCoNut & \cite{Lutellier2020cc} & \href{https://github.com/lin-tan/CoCoNut-Artifact/tree/560f5cdc79b943d167923f57f6e47c5ab3cbbb12/Results}{Github link} & All bugs in the \texttt{Results.xlsx} files are regarded correct. \\
    27 & ARJA & \cite{Yuan2020ARJA} & \href{https://github.com/TruX-DTF/APR-Efficiency/tree/6bc7e2e0a8c5a096e96f8336d62daa6900b6474a/Patches/PFL/ARJA}{Github link} & Bugs with \texttt{\_C} suffix are regarded correct. \\
    28 & GenProg & \cite{Martinez2016ASTOR,Yuan2020ARJA} & \href{https://github.com/TruX-DTF/APR-Efficiency/tree/6bc7e2e0a8c5a096e96f8336d62daa6900b6474a/Patches/PFL/GenProg-A}{GenProg-A Github link}, \href{https://github.com/TruX-DTF/APR-Efficiency/tree/6bc7e2e0a8c5a096e96f8336d62daa6900b6474a/Patches/PFL/jGenProg}{jGenProg Github link} & The results of the two GenProg tools are combined, and the date is set to the original GenProg introduction paper. Bugs with \texttt{\_C} suffix are regarded correct. \\
    29 & MutRepair & \cite{Martinez2016ASTOR} & \href{https://github.com/TruX-DTF/APR-Efficiency/tree/6bc7e2e0a8c5a096e96f8336d62daa6900b6474a/Patches/PFL/jMutRepair}{Github link} & Bugs with \texttt{\_C} suffix are regarded correct. \\
    30 & kPAR & \cite{Liu2019kPAR} & \href{https://github.com/TruX-DTF/APR-Efficiency/tree/6bc7e2e0a8c5a096e96f8336d62daa6900b6474a/Patches/PFL/kPAR}{Github link} & Bugs with \texttt{\_C} suffix are regarded correct. \\
    31 & RSRepair & \cite{Yuan2020ARJA} & \href{https://github.com/TruX-DTF/APR-Efficiency/tree/6bc7e2e0a8c5a096e96f8336d62daa6900b6474a/Patches/PFL/RSRepair-A}{Github link} & Bugs with \texttt{\_C} suffix are regarded correct. \\
    32 & Kali & \cite{Martinez2016ASTOR,Yuan2020ARJA} & \href{https://github.com/TruX-DTF/APR-Efficiency/tree/6bc7e2e0a8c5a096e96f8336d62daa6900b6474a/Patches/PFL/Kali-A}{Kali-A Github link}, \href{https://github.com/TruX-DTF/APR-Efficiency/tree/6bc7e2e0a8c5a096e96f8336d62daa6900b6474a/Patches/PFL/jKali}{jKali Github link} & The results of the two Kali tools are combined, and the date is set to the original Kali introduction paper. Bugs with \texttt{\_C} suffix are regarded correct. \\
    33 & Dynamoth & \cite{Durieux2016dynamoth} & \href{https://github.com/TruX-DTF/APR-Efficiency/tree/6bc7e2e0a8c5a096e96f8336d62daa6900b6474a/Patches/PFL/DynaMoth}{Github link} & Bugs with \texttt{\_C} suffix are regarded correct. \\
    34 & Nopol & \cite{DeMarco2014ar} & \href{https://github.com/TruX-DTF/APR-Efficiency/tree/6bc7e2e0a8c5a096e96f8336d62daa6900b6474a/Patches/PFL/Nopol}{Github link} & Bugs with \texttt{\_C} suffix are regarded correct. \\
    35 & Cardumen & \cite{Martinez2018Cardumen} & \href{https://github.com/TruX-DTF/APR-Efficiency/tree/6bc7e2e0a8c5a096e96f8336d62daa6900b6474a/Patches/PFL/Cardumem}{Github link} & Bugs with \texttt{\_C} suffix are regarded correct. \\
    36 & VFix & \cite{Xu2019VFix} & Table III of paper & Bugs with a green box are correctly fixed. \\
    37 & iFixR & \cite{Koyuncu2019IFixR} & Table 12 of paper & - \\
    38 & ConFix & \cite{Kim2019ConFix} & Table 5 of paper & All bugs mentioned in Table 5 are regarded correct. \\
    39 & Restore & \cite{Xu2020Restore} & Table 3 of paper & Rows with in which the Restore-C column's value is filled are regarded fixed. \\
    40 & LoopFix & \cite{Wang2019LoopFix} & - & We could not identify individual bugs fixed by LoopFix. \\
    41 & GenPat & \cite{Jiang2019GenPat} & \href{https://github.com/xgdsmileboy/GenPat/blob/master/repair-result/d4j-final-result.pdf}{Github pdf link} & - \\
    42 & Recoder & \cite{Zhu2021Recoder} & \href{https://github.com/pkuzqh/Recoder/tree/master/Result}{Github link} & Results collected from the \texttt{out} and \texttt{defect4j2.txt} (sic) files. \\
    43 & VarFix & \cite{Wong2021VarFix} & - & We could not identify individual bugs fixed by VarFix. \\
    \bottomrule
    \end{tabular}
    }
  \end{table*} 

\section{Language Model Architecture}
The language model uses a 5010-size vocabulary with BPE, including Start-of-Sentence (SOS) and End-of-Sentence (EOS) tokens. The Adam optimizer~\cite{Kimga2015Adam} with cross entropy loss is used for training. 
A one-layer GRU is used, with a 1000-dim token embedding layer and a hidden layer size of 1000. A single linear layer is used to project from the hidden state of the GRU to the token likelihood. 

\clearpage
\bibliographystyle{ACM-Reference-Format}
\bibliography{acmart}